\documentclass[]{sn-jnl}

\usepackage{natbib}   

\bibpunct{(}{)}{;}{a}{}{,}  
\jyear{2023}%

\begin{document}

\def\ion#1#2{{#1}\,{#2}}

\title[Science with QUVIK]{Science with a small two-band UV-photometry mission II: Observations of stars and stellar systems}

\author*[1]{\fnm{Ji\v r\'\i} \sur{Krti\v cka}}\email{krticka@physics.muni.cz}
\equalcont{These authors contributed equally to this work.}

\author[2,3]{\fnm{Jan} \sur{Ben\'{a}\v{c}ek}}
\equalcont{These authttps://www.overleaf.com/project/6414c127ff12edeec3e51f40hors contributed equally to this work.}

\author[4]{\fnm{Jan} \sur{Budaj}}
\equalcont{These authors contributed equally to this work.}

\author[5]{\fnm{Daniela} \sur{Kor\v c\'akov\'a}}
\equalcont{These authors contributed equally to this work.}

\author[6]{\fnm{Andr\'as} \sur{P\'al}}
\equalcont{These authors contributed equally to this work.}

\author[7]{\fnm{Martin} \sur{Piecka}}
\equalcont{These authors contributed equally to this work.}

\author[1]{\fnm{Miloslav} \sur{Zejda}}
\equalcont{These authors contributed equally to this work.}

\author[8]{\fnm{Volkan} \sur{Bak\i\c{s}}}

\author[5]{\fnm{Miroslav} \sur{Bro\v{z}}}

\author[9]{\fnm{Hsiang-Kuang} \sur{Chang}}

\author[1]{\fnm{Nikola} \sur{Faltov\'a}}

\author[10]{\fnm{Rudolf} \sur{G\'alis}}

\author[1]{\fnm{Daniel} \sur{Jadlovsk\'y}}

\author[1]{\fnm{Jan} \sur{Jan\'\i k}}

\author[5]{\fnm{Jan} \sur{K\'ara}}

\author[1]{\fnm{Jakub} \sur{Kol\'a\v r}}

\author[1]{\fnm{Iva} \sur{Krti\v ckov\'a}}

\author[11]{\fnm{Ji\v r\'\i} \sur{Kub\'at$^\text{11}$}}

\author[11]{\fnm{Brankica} \sur{Kub\'atov\'a$^\text{11}$}}

\author[1]{\fnm{Petr} \sur{Kurf\"urst}}

\author[1]{\fnm{Mat\'u\v s} \sur{Labaj}}

\author[5]{\fnm{Jaroslav} \sur{Merc}}

\author[1]{\fnm{Zden\v ek} \sur{Mikul\'a\v sek}}

\author[1]{\fnm{Filip} \sur{M\"unz}}

\author[1]{\fnm{Ernst} \sur{Paunzen}}

\author[12,1]{\fnm{Michal} \sur{Pri\v segen}}

\author[1]{\fnm{Tahereh} \sur{Ramezani}}

\author[1]{\fnm{Tatiana} \sur{Rievajov\'{a}}}

\author[1]{\fnm{Jakub} \sur{\v R\'\i pa}}

\author[13]{\fnm{Linda} \sur{Schmidtobreick}}

\author[1,11]{\fnm{Marek} \sur{Skarka}}

\author[1]{\fnm{Gabriel} \sur{Sz\'asz}}

\author[7]{\fnm{Werner} \sur{Weiss}}

\author[1]{\fnm{Michal} \sur{Zaja\v cek}}

\author[1]{\fnm{Norbert} \sur{Werner}}

\affil[1]{\orgdiv{Department of Theoretical Physics and Astrophysics}, \orgname{Faculty of Science, Masaryk University}, \orgaddress{\street{Kotl\'a\v rsk\'a 2}, \city{Brno}, \postcode{611 37}, \country{Czech Republic}}}

\affil[2]{\orgdiv{Institute for Physics and Astronomy}, \orgname{University of Potsdam}, \orgaddress{\street{Karl-Liebknecht-Straße 24/25}, \city{Potsdam}, \postcode{14476}, \country{Germany}}}

\affil[3]{\orgdiv{Centre for Physics and Astronomy}, \orgname{Technical University of Berlin}, \orgaddress{\street{Straße des 17. Juni 135}, \city{Berlin}, \postcode{10623}, \country{Germany}}}

\affil[4]{\orgdiv{Astronomical Institute}, \orgname{Slovak Academy of Sciences}, \orgaddress{\city{Tatransk\'a Lomnica}, \postcode{05960}, \country{Slovak Republic}}}

\affil[5]{\orgdiv{Astronomical Institute}, \orgname{Faculty of Mathematics and Physics, Charles University}, \orgaddress{\street{V Hole\v sovi\v ck\'ach 2}, \city{Praha}, \postcode{180 00}, \country{Czech Republic}}}

\affil[6]{\orgdiv{Research Centre for Astronomy and Earth Sciences}, \orgname{Konkoly Observatory}, \orgaddress{\street{Konkoly-Thege M. út 15-17}, \city{Budapest}, \postcode{H-1121}, \country{Hungary}}}
 
\affil[7]{\orgdiv{Department of Astrophysics}, \orgname{University of Vienna}, \orgaddress{\street{T\"urkenschanzstra\ss e 17}, \city{Vienna}, \postcode{1180}, \country{Austria}}}

\affil[8]{\orgdiv{Faculty of Science, Department of Space Sciences and Technologies}, \orgname{Akdeniz University}, \orgaddress{\city{Antalya}, \postcode{07058}, \country{T\"urkiye}}}

\affil[9]{\orgdiv{Institute of Astronomy}, \orgname{National Tsing Hua University}, \orgaddress{\street{101 Sec. 2 Kuang-Fu Rd.}, \city{Hsinchu}, \postcode{300044}, \country{Taiwan, Republic of China}}}

\affil[10]{\orgdiv{Institute of Physics}, \orgname{Faculty of Science, P. J. \v Saf\'arik University}, \orgaddress{\street{Park Angelinum 9}, \city{Ko\v sice}, \postcode{040 01}, \country{Slovak Republic}}}

\affil[11]{\orgdiv{Astronomical Institute}, \orgname{Czech Academy of Sciences}, \orgaddress{\street{Fri\v cova 298}, \city{Ond\v rejov}, \postcode{251 65}, \country{Czech Republic}}}

\affil[12]{\orgdiv{Advanced Technologies Research Institute, Faculty of Materials Science and
Technology in Trnava}, \orgname{Slovak University of Technology in Bratislava}, \orgaddress{\street{Bottova
25}, \city{Trnava}, \postcode{917 24}, \country{Slovakia}}}

\affil[13]{\orgname{European Southern Observatory (ESO)}, \orgaddress{\street{Alonso de Cordova 3107}, \city{Vitacura, Santiago}, \country{Chile}}}


\abstract{We outline the impact of a small two-band UV-photometry satellite
mission on the field of stellar physics, magnetospheres of stars, binaries,
stellar clusters, interstellar matter, and exoplanets. On specific examples of
different types of stars and stellar systems, we discuss particular requirements
for such a satellite mission in terms of specific mission
parameters such as bandpass, precision, cadence, and mission duration. We show
that such a mission may provide crucial data not only for hot stars that emit
most of their light in UV, but also for cool stars, where UV traces their
activity. This is important, for instance, for exoplanetary studies, because the
level of stellar activity influences habitability. While the main asset of the
two-band UV mission rests in time-domain astronomy, an example of open clusters
proves that such a mission would be important also for the study of stellar
populations. Properties of the interstellar dust are best explored when
combining optical and IR information with observations in UV.

It is well known that dust absorbs UV radiation efficiently. Consequently, we
outline how such a UV mission can be used to detect eclipses of sufficiently hot
stars by various dusty objects and study disks, rings, clouds, disintegrating
exoplanets or exoasteroids. Furthermore, UV radiation can be used to study the
cooling of neutron stars providing information about the extreme states of
matter in the interiors of neutron stars and used for mapping heated spots on
their surfaces.}


\maketitle

\section{Introduction}

The new discoveries in astrophysics during the last few decades were frequently
connected with the opening of new observational windows into invisible parts of
the spectrum. Recently, the advent of observatories working outside the
electromagnetic domain founded a new branch of science called multimessenger
astronomy. From this, it may seem that new observations in the most classical
domain of astronomy, the optical domain, can hardly revolutionize the field of
astrophysics. And yet, small and medium-sized satellite missions such as
\textit{CoRoT} \citep{corot}, \textit{Kepler} \citep{borucki2010}, \textit{MOST}
\citep{mostsat}, \textit{BRITE} \citep{britepasp}, and \textit{TESS}
\citep{ricker2015} opened a new window into time-domain astronomy, extending far
beyond the physics of variable stars and exoplanetary systems.

Similar advances may be expected from photometric missions working in other
domains of the electromagnetic spectrum. In this paper, we outline the impact of a small
two-band ultraviolet (UV) photometric satellite mission on the field of stars,
binaries, stellar clusters, interstellar matter, and exoplanets. The paper is
based on QUVIK -- Quick Ultra VIolet Kilonova surveyor \citep[hereafter Paper
I]{quvik1ssrv}, an approved Czech national science-technology mission. QUVIK is
a small 130\,kg satellite that will be launched to a low-Earth,
Sun-synchronous, dawn-dusk orbit at the end of this decade. Its telescope
will have a moderately large field of view of about $1.0^\circ\times1.0^\circ$
and two UV bands. The far-UV band is planned to have a bandwidth of
approximately 140--180 nm, and the near-UV band will span 260--360 nm. The
primary objective of QUVIK is the study of electromagnetic counterparts of
gravitational wave sources, such as kilonovae (Paper I). Besides stellar
astrophysics, the proposed science includes the study of galactic nuclei and
nuclear transients described in Paper III of the series \citep{quvik3ssrv}.

\emph{QUVIK} will provide absolutely calibrated data. This requires
calibration on the ground before the launch and onboard during the flight. The
pre-flight ground calibration procedures include spectral measurements of the
throughput of the individual parts of the optical system. The spectral response
and zero point will be determined by observing standard photometric sources
such as individual stars, white dwarfs, and open clusters. 

\section{Physics of Stars}%
\label{sec:stars}

\noindent The most important characteristic that describes the observational appearance of stars is their effective temperature because it mostly determines the stellar spectral energy
distribution.  Therefore, the effective temperature tells us what we can learn from observations of stars in different spectral domains. According to their effective temperature, stars are conventionally divided into two large groups: cool stars and hot stars.  While hot stars (with effective temperatures higher than about 7000~K) emit most of their light in UV, and therefore UV enables us to understand the physics of these stars, cool stars emit only a small part of their flux in the UV domain, most notably from outer layers of their atmospheres called chromospheres \citep{softxray}. Consequently, in cool stars, the UV tells us a lot about their activity and their impact on the surrounding environment.

Although our knowledge of stars has advanced enormously during the last century, many questions related to the evolution of stars and their impact on the surrounding environment remain unanswered.  This has a strong impact on other fields of astrophysics, including the physics of binary compact star merger sources of gravitational waves or the study of the interaction of stars and their exoplanets.  Surprisingly, even a small-size UV satellite can help to solve many of these open questions.  In the following, we explain how.

\subsection{Tests of opacities}
\label{kaptestop}

Opacities are one of the key ingredients of any numerical stellar model.  
Therefore, any uncertainty in opacity has a strong influence on the reliability
of stellar models.  For instance, inaccurate opacities are one of the suspects of the difference between the sound speed determined for our Sun from helioseismology and the sound speed derived from solar models \citep{basu}.  Incorrect opacities due to rare-earth elements may cause the mismatch between the predicted and observed light curves of kilonovae \citep{wunova}.  The inclusion of iron peak opacities into models provides a textbook example of the importance of opacities for triggering pulsations \citep{spoustazelez}. 

\begin{figure}[t]
    \includegraphics[width=0.355\textwidth]{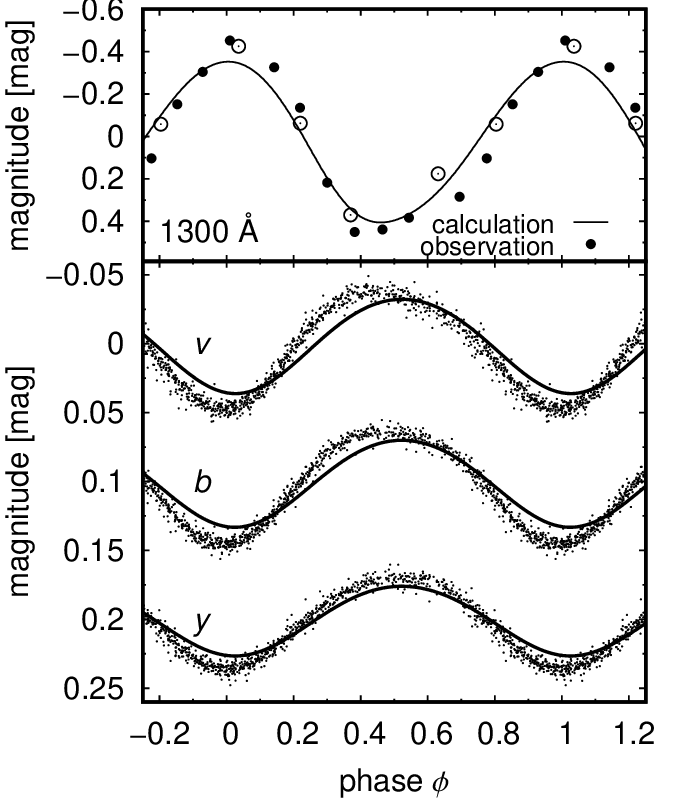}
    \includegraphics[width=0.29\textwidth]{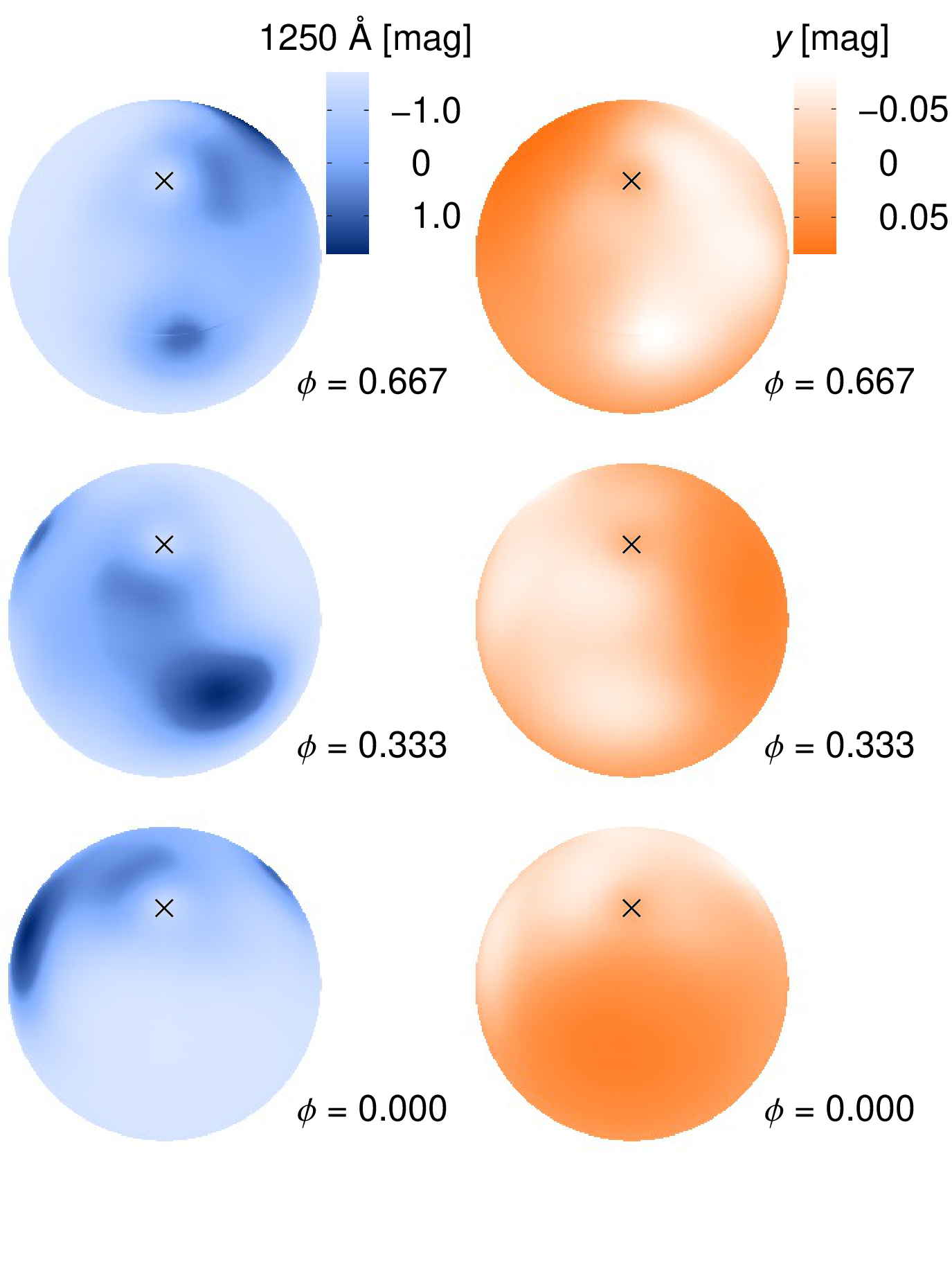}
    \includegraphics[width=0.34\textwidth]{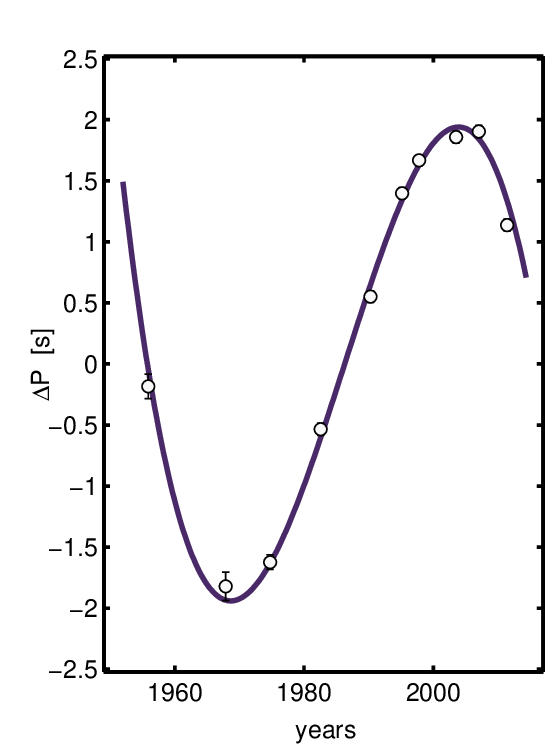}
    \caption{Variability of chemically peculiar CU Vir. \emph{Left:} The light
    curve in ultraviolet (upper panel) and visual (lower panel -- Str\"omgren
    $vby$ filters) domains.  Empty circles denote HST observations
    \citep{mycuvirhst}. \emph{Middle}: The emergent intensity at various
    rotational phases $\phi$ in the ultraviolet (blue plots) and visual (orange
    plots) domains. \emph{Right}: Rotational period variations with respect
    to the mean period derived from photometry \citep{mik2}.}%
    \label{fig:cuvir}
\end{figure}

Therefore, opacity calculations are of enormous value, and the measurement of
opacities belongs to very important tasks for laboratory astrophysics.  However,
testing opacities for actual stellar conditions  would be even more
valuable.  Here comes the hardly replaceable role of chemically peculiar stars.
These hot stars show patches with peculiar abundances of different elements on
their surfaces \citep[see Fig.~\ref{fig:cuvir}]{kochrjab}, resulting from a
complex interplay between radiative and gravitational acceleration 
\citep[e.g.,][]{2011A&A...529A..60M}.  Chemically peculiar stars show rotationally modulated light variability, caused by horizontal
variations of opacity in surface patches, which redistribute flux mostly from
far-UV to near-UV and visual domains \citep{peter, lanko}.  A comparison of
predicted and observed light curves enables us to test the opacities involved in
the modelling of the outer layers of these stars \citep[Fig.~\ref{fig:cuvir}]{prvalis, mycuvirhst}. Cool chemically peculiar stars are of special interest, because in these stars the opacities of rare-earth elements are supposed to play a decisive role.  

Because most well-studied chemically peculiar stars are relatively bright, 
it is recommended that the telescope should enable to observe also stars with
magnitude of at least 4~mag or even brighter. The typical rotational period of these stars
is on the order of days; therefore, to sufficiently cover the light curve of
these stars, the highest required cadence is 15~min for stars with the shortest
rotation period of about half of a day.  The study would benefit from
observations in the UV, where the variability is the strongest and
where antiphase behaviour of the light curves (with respect to the optical
domain) is expected.  From optical amplitudes of up to 0.1~mag,
the required signal-to-noise ratio (SNR) of individual photometric data is about
100.  For a reliable comparison with theoretical light curves, detailed
knowledge of the instrumental response function is needed.

A far-UV band enables to trace opacities due to silicon and iron, which have a
maximum in this region \citep{prvalis}.  Moreover, this band is ideal for
studying hot chemically peculiar
stars, in which the flux variability originates in the far-UV region.

\subsection{Probing the surface structure in normal B and A-type stars}

High-precision satellite photometry of normal main-sequence B and A-type stars revealed an unexpected phenomenon: a plethora of weak surface spots \citep{balonek2,balonek3}.  This discovery came as a complete surprise as hot stars lack deep subsurface convective zones. As a result, a revision of the physics of stellar envelopes comprising cool spots has been suggested \citep{balonek3}.  A more conservative approach advocates a general model of a surface consisting of overabundant spots, irrespective of the peculiar nature of the stars \citep{mymos}.

UV photometry can provide an answer about the structure of the stellar surface
in normal B and A-type stars.  Signatures of abundance spots in a spectral energy distribution differ from those of temperature spots.  Consequently, precise observation in two bands should provide data that can be tested against models, including abundance and temperature spots.

This task requires observations with an SNR of 1000 (of individual photometric
measurements) to detect the variability \citep{balonek2}.  Based on available
optical light curves, a higher cadence of observations of 5~min is needed to
obtain well-sampled light curves.  The test requires observations in a far-UV
and near-UV bands to get simultaneous observations in two bands. This combination would reliably distinguish between the two models because the expected amplitude of variability is larger in the far-UV band.

\subsection{Understanding the cyclical chromospheric activity and its evolutionary changes in cool stars}
\label{kapstarchrom}

Dynamo action powered by convection and differential rotation in envelopes of
cool stars gives rise to variable surface magnetic fields and activity
\citep{chatar}.  A strong magnetic field suppresses convective energy transport
in localized surface regions, leading to the formation of cool surface spots,
which appear dark due to their lower effective temperature \citep[e.g.][] {granz}.  On top of that, part of the energy flux transported by subsurface convective motions heats the chromospheres and coronae of
cool stars.  Except for variations due to flares, the UV flux shows rotational variations \citep{halvlk}, variations during the activity cycle as measured in our Sun using the SORCE satellite \citep{sorce}, and variability in the course of the stellar lifetime \citep{gui}.

The evolutionary variability of the UV flux, that is, change of both mean
flux and the properties of short-term variations,  can be determined from
observations of individual cool stars with different masses and ages.  This can
tell us if the UV chromospheric flux varies with age in a similar way to X-ray
activity or rotation \citep{ryby}. Such a study requires observations below
285~nm, where the chromosphere contribution dominates \citep{softxray,
celysten,johnlutter}. Therefore, far-UV band of \emph{QUVIK} could be used to
probe the chromosphere, while the near-UV band would be contaminated by
photospheric flux. Alternatively, simultaneous UV observations in the
220--290~nm band with \textit{ULTRASAT} \citep{ultrasat} would provide a
complementary probe of the chromosphere. The far-UV band is also important for
the interaction of cool stars with their exoplanets (see Sect.~\ref{secexoact}).
The expected evolutionary variability is relatively large, therefore, a
medium SNR of about 100 is required. 

Besides the evolutionary variability, cyclical variability of chromospheric
emission may also have an impact on planetary atmospheres on the timescale of
several orbits.  While the basic period of solar activity is typically too
long to be detected by a single small satellite, other stars show shorter
periods of activity of a few years \citep{wilsonchrom, nedal}. Such a
periodicity can be detected even by a mission with a duration shorter than the
solar cycle. As a result of similar amplitude of flux changes, the
requirements of such observations are basically the same as for the study of
evolutionary changes, except for cadence, which is required of about 14
days due to an expected order-of-year-long period of this type of
variability.

The flaring activity of our Sun became proverbial.  However, flares are more
typical in stars much cooler than the Sun, in M dwarfs, where the star may
brighten by several orders of magnitude \citep{froning}.  Flares are the
aftermaths of a sudden release of magnetic energy that affects the stellar
surface \citep{hai}.  This locally heats the atmospheres to temperatures of the
order of tens of thousand kelvins and leads to a brightening of a star
\citep{celacervena}.  Due to the high temperature of flaring matter, flares can
be best studied in the ultraviolet domain \citep{profesional, flargal}.  The
study of UV light curves of flaring stars can help to understand the properties
of flares \citep{froning}, their frequency, and their influence on surrounding
exoplanets. To fit the light curves of flares and to determine flare properties
\citep{kowall}, observations in both far-UV and near-UV bands
are required.  Because the flares can appear on a timescale of seconds and their occurrence cannot be predicted, monitoring of a given star for days with a cadence of 1~s is needed to
determine the flaring activity. A medium SNR of 100 is sufficient for this purpose.

\subsection{Nature of warps in the light curves of magnetic stars}

\begin{figure}[t]
    \centering
    \includegraphics[width=0.5\textwidth]{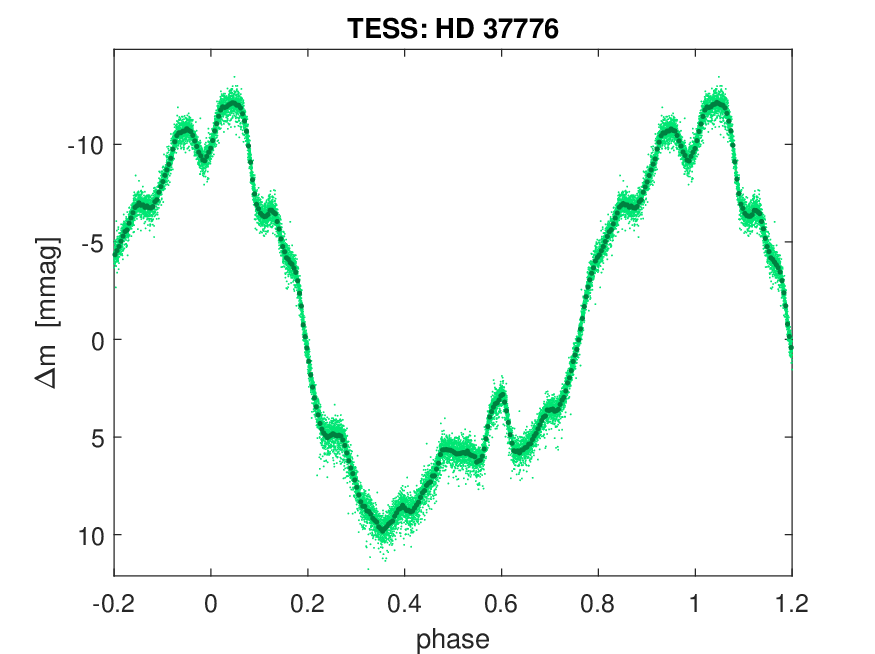}
    \caption{TESS light curve of a chemically peculiar star HD 37776.  Besides
    the large-amplitude rotational variability (due to horizontal opacity
    variations), the light curve shows regular short-term variability dubbed
    `warps' by \citet{mikland}.}
    \label{fig:mikland}
\end{figure}

High-precision satellite observations of chemically peculiar stars revealed
multiple persistent phase-locked warps on their light curves \citep[see
Fig.~\ref{fig:mikland}]{mikland}.  The warps likely originate due to the light
absorption in centrifugally supported magnetospheric clouds reminiscent of
stellar prominences \citep{mraky}.  The clouds are expected to be fed by wind,
providing the only evidence of winds in low-luminosity hot stars.
However, the competing model predicting the other origin of the warps due to
surface features was not yet excluded.  High-precision observations in two bands
can resolve the issue.  Magnetospheric absorption originates on free electrons;
therefore, it is predicted to be grey.  On the other hand, variability due to
surface features is expected to show different amplitudes in different bands.
Therefore, precise photometric observations in two filters will allow us to
understand the nature of warps in the light curves of chemically peculiar
stars and possibly prove the existence of stellar wind. Observations with a relatively high SNR of 1000 are required to clearly determine the amplitude of the warps (see Fig.~\ref{fig:mikland}).  The corresponding cadence of observations should be
about 1~min to observe warps with high time resolution. The near-UV band combined with optical photometry is suitable to distinguish between the models of variability. Alternatively, a far-UV band would enable an even better test of the nature of light absorption in hot stars with corotating magnetospheres \citep{mysigorie} and extremely hot white dwarfs \citep{uhlprom}.

\subsection{Determining the drivers of line-driven wind instability} 

Photometers on board satellite missions detected low-amplitude light variations
of single O-type stars.  These variations were attributed to gravity waves
\citep{aeovar} or subsurface convection \citep{blomcor}.  However, there is an
intriguing possibility that this variability bears the signature of a
line-driven wind instability \citep[deshadowing instability,][]{oblavar}.  This would imply that we are able to detect seeds of line-driven wind instability in photometry.  Understanding and detailed modelling of this instability are crucial for deriving the precise wind mass-loss rate of hot stars from observations \citep[e.g.][]{chuchcar, clres2}. 

Continuous photometric observation of selected stars in two bands is needed to
distinguish between individual sources of stochastic light variability. Wind
blanketing and changes of temperature affect atmospheres in different ways.
Therefore, the amplitudes and signatures of the variability in two bands provide
the key observables in this case \citep{oblavar}. This requires continuous
photometry for several days.  A relatively high SNR of 1000 and cadence of about
30~s is needed to clearly detect the variability. A reasonable option is to use
a near-UV band in combination with a far-UV band, where the light variability due to wind blanketing is strong  \citep{oblavar}.

\subsection{Unveiling the mass-loss histories of Be and B[e] stars}

Massive stars are born with masses higher than about 8 M$_\odot$, but leave a
compact remnant with a typical mass of a few solar masses.  When did the mass
get lost? We are not sure \citep{2014ARA&A..52..487S}, and that is why we have
to explore different evolutionary stages to understand when mass loss happens.
B[e] stars and most notably B[e] supergiants belong to the evolutionary stage
during which a significant amount of mass can be lost.  Long-term UV
observations of these stars can distinguish between individual states of the
envelope caused by, e.g., variable disk absorption or LBV-type variations
\citep{hutse, shorec, beuv}.  Furthermore, UV variability provides information
on the structure of the envelope in Be stars \citep{uvsmith}.  Observations in
two UV bands revealed that some B[e] stars maintain constant luminosity
during their photometric variations, providing a better understanding of the drivers behind
the variability \citep{beuv}. The study also showed that the UV band is
ideal for detecting pulsations, enabling us to understand the mechanism behind
the outbursts in these stars.

A typical timescale of the variability of B[e] star circumstellar environment is
of the order of months; therefore, a cadence of observations of 14 days should be
sufficient.  The variability shows relatively large amplitudes, which translates
into a loose constraint on SNR of about 10. The study requires observations in
the near-UV band, where the variability due to variable effective temperature and dust obscuration is strong. The analysis would benefit from simultaneous UV observations in the 220--290~nm band with \textit{ULTRASAT}, whose band is located close to the carbon opacity bump. The inclusion of a far-UV band would enable the construction of near reddening-free colour-magnitude diagrams \citep{beuv}, which probe luminosity variations.

Be stars display variability on various time scales.  Usually, this variability
is described as long-term (years), medium-term (days to tens of days), and short-term
(fraction of a day).  This variability is caused by different mechanisms, such as
disk growth and dissipation or disk waves \citep{rivirev} and pulsations
\citep[e.g.,][]{bebrite} and is visible in
spectral line profiles and photometry.  There exist observational records of
visual photometric observations of Be stars
(e.g.\ \citealt{kresohvar}, \citealt{keltaj}).
Simultaneous observations in visual and far-UV or near-UV bands would significantly help to understand the Be star variability.  The observational effort may include all variability time scales, from short to long. The required cadence depends on the properties of the objects.

\subsection{Mass-loss mechanism in cool supergiants}
\label{betelgeuse}

After decades of studies, the wind mechanism in cool supergiants is still not
very well understood.  Processes invoked to explain cool supergiant outflows
include the radiation force on dust grains, pulsations, and convection.
\citet{epizoda} suggested that most of the mass-loss in these stars is connected
with episodic gaseous outflows similar to that linked to the Great Dimming of
archetypal cool supergiant Betelgeuse (see also
Sect.~\ref{sec:transiting_dusty_objects}).  This can be tested using UV
observations. The proposed mechanism for this dimming, which includes dust cloud
\citep{levesque,2021sf2a.conf...13M} and changes in stellar effective
temperature \citep{dharma}, has to cope with the near constant brightness of the
star in the infrared \citep{betnedim}.  Variable dust absorption and changing effective temperature can be distinguished, especially at short wavelengths, due to the different dependence of amplitudes on wavelength \citep{jadbet}.

A typical time delay between ejections of individual clumps of the order of
years \citep{2021AJ....161...98H} is comparable to the duration of a small
mission. Therefore, the frequency of episodic gaseous outflow events in cool
supergiants should be estimated from the observation of a sample of these stars
in two UV bands, including the near-UV band and either far-UV or optical band.  Given the long timescale of the variability of these stars, the 1-day cadence should be sufficient with the required SNR ratio of about 100.

\subsection{Understanding pulsating stars of the upper main sequence}

Pulsating stars allow one to probe stellar interiors, which are not accessible to direct observations.  Therefore, the physical conditions and processes within a star can only be observed through their influence on the pulsation periods and amplitudes.  Pulsations provide an elegant way to derive stars' basic fundamental properties, such as mass, radius, and distance \citep{2009Natur.459..398D,2011Natur.471..608B,2011A&A...530A.142B,2013MNRAS.429..423M}. 

Several different types of pulsating stars appear in the upper main-sequence
region of the Hertzsprung-Russell diagram. They include $\beta$~Cephei stars,
which are very massive p-mode (pressure mode) pulsators
\citep{2022A&A...659A.142S}, $\delta$~Scuti stars, which are multiperiodic
pulsating variables located in the lower part of the Cepheid instability strip,
and rapidly oscillating Ap stars \citep{kurpreh}, which belong to chemically
peculiar stars with pulsational variability in the period range of 5 to 25
minutes. Pulsating stars of the upper main sequence include also g-mode
(gravity mode) pulsators, such as slowly pulsating B-type stars, which pulsate
in high-order modes \citep{2018A&A...618A..47C}, and $\gamma$ Doradus
stars, which are intermediate-mass, late-A/early-F stars, located  at the cool
border of the classical $\delta$ Scuti instability strip, pulsating in
high-order g-modes \citep{2000ApJ...542L..57G}.

If we look at any models describing these classes of variable stars, we find
that the amplitudes typically increase significantly with decreasing
wavelength \citep{2021FrASS...8...97G}.  The increase is strongly dependent on
the rotation (especially of the core), the metallicity, and mass-loss rate.
However, in the UV we have so far only sparse information from ground-based observations in Johnson $U$, for example.  Time-series in the UV are, therefore, very much needed to test and calibrate our current pulsational models.  This will be an excellent supplement to the data from the \textit{Kepler}, \textit{PLATO}, and \textit{TESS} missions. 

To advance the physics of stellar pulsations, simultaneous (or near-simultaneous within minutes) observations in two bands are required. Both bands should be preferably located in UV to obtain a larger amplitude of variability.  The benefits of two-color photometry are illustrated, e.g., by \citet{weissuni}. Individual pulsating stars show very different periods; therefore, depending on the variability type, the required cad ency is from 5 minutes up to days.  For the pulsational analysis, a high SNR of about 1000 is needed.

\subsection{Neutron stars}%
\label{sec:neutron_stars}

\noindent Neutron stars (NS) are strongly magnetised ($10^8$--$10^{13}$~G), 
high-density ($10^6$--$10^{14}$~g\,cm$^{-3}$) fast-rotating objects with periods from milliseconds to several hundred seconds.  They are unique laboratories to study degenerate matter inside peculiar objects \citep{Lorimer2004}.  In addition, there are also specific types of NSs that manifest further kinds of accompanying physical effects -- pulsars, accretion-powered NSs, interacting NSs in binaries, merging NSs, and magnetars (a type of NS with an extremely strong magnetic field).  Magnetars might be related to the recently discovered phenomenon of fast radio bursts (FRBs), mysterious millisecond transients currently only detected at radio frequencies whose nature is still unclear \citep{CHIME2020,Xiao2021,Petroff2022}, see Paper~I.

There is still a lot of uncertainty about the evolution of NSs, their internal
structure and the structure of their magnetospheres, the sources of their UV
emission, locations of these emission regions, and whether there is, and
to what degree, a relation between magnetars and FRBs.  UV observations are
crucial to resolve these questions, but are very scarce to date and could be relatively easily provided by a small satellite.

\subsubsection{Revealing the internal composition of neutron stars by studying their thermal evolution}

NSs cool down via photon and neutrino emissions at timescales of tens of
millions of years.  This evolution provides an important way to reveal NS
composition, properties, and equation of state.
If an NS cools only passively, its effective temperature is
expected to drop below $10^4$~K in $\sim 10$~Myr. However, various heating
processes slow down (and in some cases even reverse) the cooling
\citep{Gonzales2010}.  For old NSs, the measured effective temperatures
have been shown to be significantly higher than predicted by cooling models
\citep{Abramkin2022}.  There are models of heating processes such as
rotochemical heating and heating due to magnetic field decay. Moreover, NSs were proposed to attract dark matter because of their high mass.  As dark matter reaches the interior of the NS, its interactions with neutron matter might lead to a measurable heating of the star \citep{Kopp2022}.

Because the maximum thermal emission of NSs is in the UV range due to their high effective temperatures, these processes are difficult to be observed from the ground. To properly estimate the spectral index in the UV, observations should be made in at least two filters. Additional spectral points necessary to fit black-body radiation can be obtained from observations at visible and X-ray wavelengths from archives.

A specific objective for the small satellite can be to estimate the
effective surface temperature of five to ten middle-aged to old NSs.
By observing in two UV filters, quantitative estimates of the NS surface
temperature can be obtained and compared with the temperature from NS
evolution models.
Constraints on the NS effective temperatures would
allow refinements of the evolution models and constrain the internal structure,
composition of NSs, and NS heating process(es). Since old NSs are stable on long timescales, observations are not time critical and can be obtained at any time during the mission.

We identified seventeen candidate pulsars with expected magnitudes 
brighter
than 22~mag in the UV range \citep{Hog2000, Hobs2004, Mignani2011, Gaia2020}.
All have been found at least at visual and radio wavelengths simultaneously. If
no UV data were available, we estimated their brightness to be 2 magnitudes
fainter than the V-band  magnitude, based on the typical intensity difference of
the known pulsars. Also, this difference is consistent with old NSs that
have been simultaneously observed in the visual and UV range. Some of the
candidates are already covered in UV \citep{Abramkin2022}; however, these
measurements are not suited to estimate spectral indices, for which we would
need observations made in two UV filters. For a NS magnitude of 22~mag, expected
as a practical limit by \emph{QUVIK}, we get $\mathrm{SNR} \sim 23$ in $\sim 20$
hours, which is sufficient for the proposed estimations of the spectral slope of
the black-body radiation.

\subsubsection{Mapping the emission regions on the neutron star surface}

At the surface of the NS, heated regions can be formed, which have been observed in soft X-rays \citep{Guillot2019}.  These are magnetic pole regions where relativistic particles (the so-called returned current) from the magnetosphere can hit and heat these spots.  During the rotation of the NS, the measured flux changes due to the movement of these heated regions, allowing it to constrain the impact of individual regions.  
However, the observed UV radiation may originate from these regions (hot spots) with a thermal nature or from other locations in the NS magnetosphere with a non-thermal nature, or a combination of them. 
For the magnetospheric origin, the emission mechanisms and locations remain unclear. More recent models for the non-thermal emission from pulsars include, for example, the striped-wind model \citep{2015A&A...574A..51P}, the extended-slot-gap and equatorial-current-sheet model \citep{2022ApJ...925..184B}, the synchro-curvature model \citep{2022MNRAS.516.2475I}, and the non-stationary outer-gap model \citep{2017ApJ...834....4T}. 
UV observations to disentangle the thermal and non-thermal components can help to constrain theoretical models.
Moreover, the emission mechanisms and regions of magnetars are very likely different from those of other classical neutron stars.   If the heated surface regions are located where the magnetic field penetrates the star, monitoring of the emission regions allows refining the long-standing and highly-discussed issue between a dipolar (classical NS) or multipolar (magnetars) shape of the magnetosphere, the relation between the different magnetosphere shapes and their impact on the NS evolution \citep{Yao2018}.

UV photometry can provide an important way to model the thermal spots and shape of the magnetosphere.  To characterise the location of their emission regions, the aim can be to determine the UV light curves for pulsars (periods $< 1$~s) and magnetars (periods $\sim 10$~s) in the near- and far-UV filters, compare the light curves with those observed in X-rays and visual range and localise the emission regions based on inverse modelling \citep{Riley2019}.

The candidate objects have $\lesssim 16$~mag for pulsars and $\lesssim18$~mag
for magnetars in the UV range.  Each measurement should be long enough to obtain
$\mathrm{SNR} \sim 70$ for pulsars and $\mathrm{SNR} \sim 45$ for magnetars.
The exposure time can be divided into sub-exposures, each of them set to $\sim
1/15$--$1/30$ of the NS rotation period (typically 1~ms--1~s). The sub-exposures
have to be aggregated according to the NS rotation phase to obtain the average
phase curve.  This short-time observing mode would be unique for the UV
observations of a small satellite and can only be provided by a very few
instruments.  The observations can be obtained over the whole mission in more
time intervals, given that the precision of the sub-exposure time stamp is
significantly higher than the exposure time itself.  Furthermore, as some of the
NSs show sudden changes in their rotation phase (called glitches), their
measurements have to be done in a time interval of a day or a few days.  The
fast read-out window of the chip has to be scalable and large enough to include the whole PSF of the stars. Similarly, the exposure time has to be changeable down to a few milliseconds.

\section{Binary stars}%
\label{sec:binaries}

A substantial fraction of stars are found in binary or multiple stellar systems.
The binary star fraction depends on the stars' spectral types
\citep{moedistefano}.  Most massive and luminous stars appear typically in
binaries or triple systems; on the other hand, red dwarfs spend their lives
mostly alone.  Binary stars are crucial in astrophysical research because they
are unique sources of information.  Their orbits provide the most important
stellar parameter, the masses of stars.  The most precise and direct
determinations of masses are enabled using visual or eclipsing binaries
which are also spectroscopic binaries \citep{serenelli}.  The values of masses
at an accuracy better than 1--3\%  are necessary for tests of models of stellar
evolution \citep{andersen,torres}.  Binary stars provide an extremely large
amount of data about stars, their atmospheres, stellar winds, plasma
physics, and the interaction between stars and their surroundings. They can be
used for an independent distance determination \citep{southworth,graczyk}.

The world of binaries is very diverse.  In detached pairs in relatively wide
orbits, the companions evolve almost independently, allowing us to accurately
determine the component masses, radii, and temperatures. By taking into account
their age and chemical compositions, we can use them to constrain and calibrate
models of stellar evolution.  Such systems were used to study the primordial
helium abundance, the helium-to-metals enrichment ratio \citep{paczynski,
brogaard}, the convective core overshooting, the mixing length parameter
\citep{guinan,claret}, mass loss by stellar winds \citep{ignace}, and standard
luminosities and multiband bolometric corrections \citep{bakis22}.  If the stars
in a binary system are close enough to fill their Roche lobes, we speak about
semi-detached or contact binary.  In such close binaries, mass exchange between
companions or mass loss from the system occurs during their lifetime.  Many
different phenomena can then be studied: tidal effects, mass transfer,
mass loss, and orbital angular momentum loss (e.g.\ \citealt{loukaidou}).  When
the mass transfer occurs during the supergiant evolutionary phase of the more
massive component, the star is stripped off its envelope. Consequently, the
evolution may lead to objects such as cataclysmic variables, X-ray binaries,
(super-, \mbox{hyper-,} kilo-)novae, millisecond pulsars, double-degenerate systems,
and generation of short gamma-ray bursts or gravitational waves \citep{podsiad,
anton}.

\subsection{Galactic and extragalactic massive binaries}

The origin and further evolution of massive binaries are still not fully
understood. However, these binaries are of special interest, because the final
stage of their evolution results in very bright transient events caused by
super-, hyper-, kilonovae, or mergers generating gravitational waves. To
evaluate and improve the models of stellar evolution, we need to know the masses
and radii of stars.  The easiest way to obtain these parameters is to study
eclipsing double-lined spectroscopic binaries.  There are more than two million
known eclipsing binaries \citep{gaia}.  However, only less than 300 have the
fundamental parameters determined with accuracy in 1--3\% as requested for the
test of stellar models \citep[see DebCat catalogue in][]{debcat} and only about
one-fourth of the included systems have massive components (over $3\, M_\odot$).
Thus, new precise measurements in the bands, where the hot massive stars radiate
predominantly, are highly demanding.  Observations in UV and visible bands help
to determine the radii of binary components. Two-channel observations in UV
could increase the accuracy of determination of the effective temperature ratio
and allow us to model the reflection effect \citep{pigulweb}.

To provide precise binary parameters, accurate observations in two 
bandpasses with a time resolution of less than 5 minutes covering the whole
phased light curve of selected systems are required. A combination of 
near-UV and the optical band could be used, but near-UV and far-UV
provide better constraints for the binary system model.

\subsection{(Post-)common envelope binaries}

During their lives, a large number of binaries undergo a phase in which both stars share one common envelope.  Such evolutionary phase and resulting post-common-envelope binaries (PCEBs) are predecessors to many extraordinary objects such as SN\,Ia supernovae, X-ray binaries, double degenerated binaries and others \citep{paczynski76, wang, ivanova}. 

Nowadays, surveys provide valuable data for searching PCEBs.  The number of 
such systems containing late-type stars has increased rapidly. 
However, white dwarfs in PCEBs with the early-type component are outshined by
their companions.  The radiation of white dwarfs only makes up 1\% of the total
radiation of binary.  However, the use of a combination of far-UV, 
near-UV, and optical data facilitates the detection of these PCEBs.  A
convenient way to detect such systems is to search for main-sequence stars with
substantial UV excess as a result of the presence of a nearby white dwarf
companion. This will allow us to determine the occurrence rate of PCEBs with
early-type components and search for this kind of progenitors of exploding stars
or mergers.

This program could be executed as complementary and use observations of binaries
that, by coincidence, appear in field of view (FoV) while monitoring other primary targets.

\subsection{Symbiotic binaries} 
\label{symbiotics}

Symbiotic variables are strongly interacting binary systems consisting of a cool evolved star (typically M or K giant) and a hot component (white dwarf or neutron star). These objects are unique astrophysical laboratories in the study of stellar evolution, mass transfer, accretion processes, stellar winds, jets, dust formation, or thermonuclear outbursts \citep[e.g.][]{2012BaltA..21....5M,2019arXiv190901389M}.

The symbiotic stars are divided into two categories: so-called shell-burning
symbiotic binaries with hydrogen burning ongoing on the surface of the white
dwarf and accreting-only symbiotic stars powered by accretion. The latter group
is less represented in the catalogues of symbiotic binaries
\citep[e.g.][]{2000A&AS..146..407B,2019RNAAS...3...28M,2019ApJS..240...21A} as
the optical spectra of these symbiotic stars are often inconspicuous. They can
be detected due to UV excesses, significant flickering (which is most prominent
in near-UV), or their X-ray emission 
\citep[e.g.][]{2013A&A...559A...6L,2016MNRAS.461L...1M,2021MNRAS.505.6121M,2023arXiv231216126M}. 

UV photometric mission can search for new symbiotic binaries via detection of
the UV excess of the red giant branch and AGB stars. UV photometry can
also be used for verification of the symbiotic candidates by searching for
UV excess and/or UV flickering of the candidates. Such studies require just one
UV band and 30--60 minutes time-series of short exposures of a few tens of
seconds to search for flickering. Time series of UV observations can also help
to study the process of accretion in symbiotic stars and in probing the physical
mechanisms of the classical symbiotic outbursts, which is still debated
\citep{2019arXiv190901389M}. In this case, UV observations of symbiotic stars
performed in two bands supplemented with ground- and space-based optical data
can put constraints on the physical mechanisms of the outbursts. Together with
optical and infrared (IR) spectral energy distribution, photometry in two
UV bands can also be used to improve parameters of the hot components of known
symbiotic binaries. Long-term light curves (2--3 years) in one UV band sampled
with a cadence of a few weeks can provide the orbital periods of symbiotic
binaries on the basis of the reflection effect, which has the highest amplitude
in near-UV.

\subsection{Cataclysmic variable stars}

Cataclysmic variables (CVs) are close, interacting binary systems comprising a
white dwarf (WD) receiving mass from a Roche lobe-filling late-type star.  The
actual accretion process is determined by the absence or strength of the WD's
magnetic field. For non-magnetic CVs, an accretion disc forms around the WD,
which helps to remove the angular momentum of the in-falling material. The gas
flows and connected processes in CVs are very complex.  The WD and the accretion
disk reach temperatures up to $10^4$--$10^5$~K; thus, their behaviour is
observable best in UV\@.  In fact, the total luminosity of an accreting CV is
dominated by the luminosity of the accretion disk, which can be derived from the
absolute UV magnitudes \citep{selvelli_gilmozzi_2013}. 

The evolution of a CV is driven by the loss of angular momentum.  This process
controls the change of the orbital period and, in the phases of interaction, the
mass transfer rate.  For systems above the period gap, magnetic braking
dominates the angular momentum loss, while for systems below the gap,
gravitational radiation is the main contributor \citep[for a summary,
see][]{knigge+2011}. The total luminosity gives the momentary accretion rate;
for dwarf novae, differences are observed between quiescent states (no
accretion) and outbursts when the material from the disc is dumped onto the
white dwarf. One of the key questions that can be addressed by a UV photometry
mission is how the momentary mass-loss rates determined from the total
luminosity compare to the secular accretion rates as revealed by WD temperature
or secondary bloating. Time-domain UV observations could also help to probe the
variations of the accretion rate with the orbital period. This can provide tests
of theoretical models of angular momentum loss.

Comparison of these accretion rates with standard models shows that there is a
possible increase in the accretion rate in the orbital period range of 3--4~h,
i.e., \ just above the period gap.  This is the domain of SW~Sex stars, an
evolutionary state that CVs seem to go through before they enter the 
period gap \citep{schmidtobreick2013}.  While the SW~Sex phenomena can be
explained by assuming high accretion rates, actual values for the accretion rate
have only been estimated for three of these objects \citep[][and references
herein]{2017MNRAS.466.2855P}.  Measuring the accretion
rate from the UV luminosity and comparing it for SW~Sex stars, classical 
nova-like stars, and dwarf novae will help to understand these extreme objects. 

Classical novae are CVs that have been observed to have a nova eruption.  Theory
predicts that the WD is heated by the eruption.  Subsequently, it can irradiate
the secondary and thus influence the accretion rate.  This would result in a
change in luminosity. \citet{selvelli_gilmozzi_2013} compared the luminosity and
spectral energy distributions of several novae at different epochs using
data from various UV satellites.  For most systems, they did not find a
significant change over the time of ten to twenty years.  A comparison with new
data taken few decades later would indicate if any variation in the luminosity
of the accretion disc is present on larger time scales.

For the studied dwarf novae, the analysis requires to measure the brightness
during several outbursts. This will enable to determine the total luminosity and
accretion rate for each measurement; from this, the overall accretion during a
cycle can be established and then calculated into an average accretion rate for
the system. For nova-like stars (normal ones and SW~Sex type) and old
novae, snapshot observations in two UV bands and are sufficient as the accretion
discs are stable, and the average accretion rate can be derived directly.

\subsection{RS CVn-type stars} 

RS CVn-type stars are variable stars found in detached binary systems.  They are
composed of main-sequence stars with spectral types around G--K\@.
\citet{1986ApJS...60..551F} gave three properties required to become a member of
this group of variable stars: at least one star must exhibit emission in the H
and K resonance lines of \ion{Ca}{II}, which are created in the 
chromospheres of stars, systems must show light variations caused by 
effects other than eclipses, pulsations, or ellipticity.  The more active star
must be evolved, and it also has to have spectral type F, G, or K\@.  

Brightness variations of RS CVn-type systems are caused by the large stellar
spots.  Typical brightness fluctuation is around 0.2\,mag.  The spots are
created by high chromospheric activity, similar to the Sun.  Some RS CVn-type
systems exhibit brightness variations due to stellar eclipses superimposed 
on changes caused by the variation in the spot surface coverage fraction, e.g.\
RT~And, XY~UMa \citep{2000A&A...362..169P, 2001A&A...371..997P}.  Photometric
variability is often correlated with chromospheric line emissions.  UV emission
is, by solar analogy, known to be associated with stellar active and transition
regions (Sect.~\ref{kapstarchrom}).  These areas on the Sun are associated with
intense magnetic fields, and sunspot activity is enhanced in and around these
magnetically active regions.  Such magnetic activity is also detected in RS
CVn-type systems. \citet{1998ESASP.413...91M}, using combined optical and UV
observations, showed a spatial correlation between photospheric spots and
chromospheric plages.

Therefore, photometric observations with a UV satellite could enable studying
evolution of energy loops and magnetic activity, their impact on the formation
of active regions, and stellar spots. Such analysis requires multi-band
observations of the whole-phased light curves of selected objects with a time
resolution of up to 10 minutes.

\subsection{Reflection effect in binary systems}

Reflection effect is crucial
in close binaries with extreme temperature differences between the components.  It is also important in connection with exoplanets since some of them (e.g.\ hot Jupiters) can be considered extreme cases of binary stars subject to extreme reflection effects.  The reflection effect describes the mutual irradiation of the components in a binary system and, most importantly, the surface temperatures of the components and the radiation field.  Because one of the components is
typically
relatively hot and temperatures on the irradiated side of the cool component can exceed 10\,000~K, UV observations are key to our understanding of this effect.

There are models of the reflection effect with various degrees of sophistication and precision.  Models that are used to calculate the light curves and spectra of eclipsing binaries are the most popular \citep{wilson90}.  They usually assume that the irradiated atmospheres radiate as non-irradiated ones, but with an enhanced effective temperature.  Apart from that, some models take into account a parameterisation of heat redistribution over the surface and scattered light \citep{budaj11,horvat19}. 
There are
LTE and NLTE models of stationary irradiated atmospheres in radiative and convective equilibrium \citep{burrows08,vuckovic16} 
revealing
that irradiated atmospheres may have a very complicated structure with
temperature inversions that can lead to the appearance of emission lines.
Finally, there are hydrodynamic simulations of the process \citep{dobbs08}
indicating the presence of an equatorial jet carrying the heat from the 
irradiated to shadowed side.

Despite all these efforts on several fronts, there are plenty of open questions
that need to be addressed, and UV observations of binaries with one hot
(WD, sdB, sdO) and one cold component (a red dwarf or a red giant) will play a crucial role. 
For example:
What is the amount of scattered light and albedo of the irradiated star?
Why do some observed albedos of irradiated stars require values close to or larger than one?
What is the efficiency of the heat transport from irradiated to
nonirradiated side in (binary) stars/exoplanets? 
What is the structure and radiation field of the irradiated atmosphere?
How does it affect the albedo, convection, shape of the star, gravity, and limb darkening \citep{rucinski69, claret11}?
How does the reflection effect affect stellar and planetary evolution?
To answer these questions, multi-band observations with an SNR of about
100 with a time resolution of at least 5 minutes are needed.

\subsection{Eclipsing binaries: a key to deriving bolometric corrections in the UV}

Recently, eclipsing binaries with well-known physical parameters of the
components were shown to be a very important tool to derive multi-band
(Johnson $B, V$ and Gaia $G, G_{BP}, G_{RP}$) bolometric corrections
\citep{bakis22}, which can be used to derive luminosities of the
components with uncertainty  as low as one percent. \cite{eker23}
increased the number of photometric bands in their method from 5 to 6 by
including the \textit{TESS} \citep{TESS2014,ricker2015} bandpass to derive
bolometric corrections for the \textit{TESS} bandpass. They also
showed that it is possible to derive luminosities and radii of single
stars with an error of about 2 per cent, which is equivalent to values
already published. Thus, increasing the number of bandpasses to achieve
more reliable luminosity and radius of binary components and single stars is
paramount. The key issue of this method is to obtain precise magnitudes in the
desired bandpasses. It is obvious that using magnitudes of stars obtained
in UV, which are only possible by systematic space-based  observations of
binary systems and/or single stars, will fill the gap on the UV side of the
wavelength regime of the method described by \cite{bakis22}.

\section{Mergers of intermediate masses}\label{sec1}

N-body simulations of \cite{Nela-Nbody} show that about 50\% of the merger
products of the binaries are B-type stars. The first post-merger in the phase
when the envelope started to be transparent was found among FS~CMa stars (IRAS
17449+2320, \citealt{IRAS_AA_mag}). FS~CMa stars are a~subgroup of stars showing
the B[e] phenomenon, i.e., the presence of the forbidden emission lines and
infrared excess \citep{Allen76,Lamers98}. These are indicators of a very
extended circumstellar gas and dust region. Properties of several other FS~CMa
stars indicate that there may be more post-mergers hidden in this group.
However, not every object in the list of FS CMa stars is a~post-merger. The
group was defined primarily on the basis of the infrared properties
\citep{Sheikina_warm_dust_00,Miroshnichenko_warm_dust_01,
Miroshnichenko_FS_CMa}, which is not a~sufficiently sharp criterion. Therefore,
the list contains several binaries (Be/X-ray binary CI~Cam,
\citealt{Barsukova06AR}; 3~Pup, \citealt{Miroshnichenko20_3Pup}; GG~Car,
\citealt{Porter21}; AS~386, \citealt{Khokhlov18}; or MWC~349A,
\citealt{Tafoya04}),  B[e] supergiant (MWC~300, \citealt{Appenzeller77}), and
post-AGB candidates (Hen3-938, \citealt{Miroshnichenko99_Herbig}, CPD-48 5215,
\citealt{Gauba04}).

\begin{figure}[h]
\centering
\includegraphics[width=0.9\textwidth]{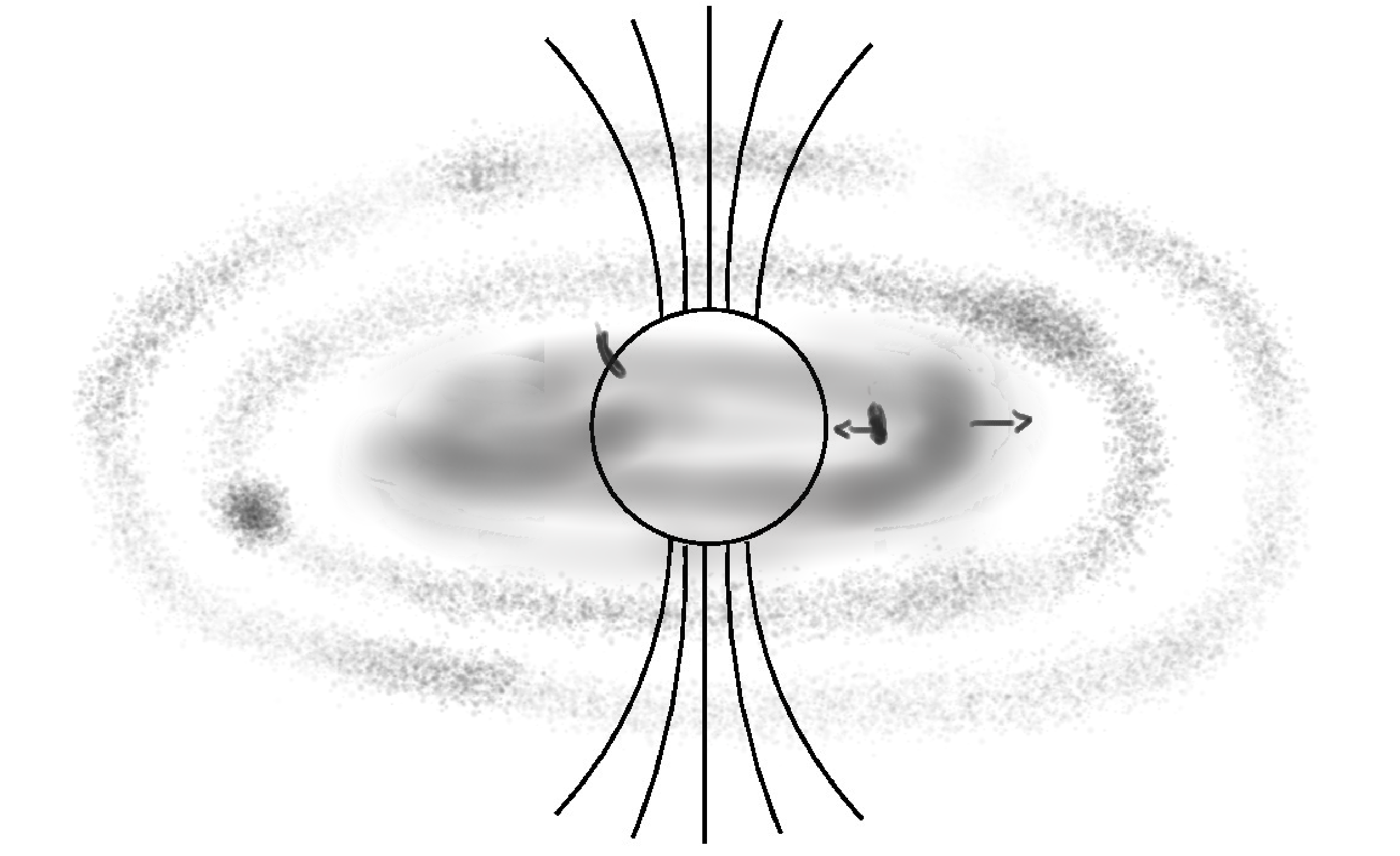}
\caption{Empirical model of a FS~CMa post-merger. The slowly rotating strongly magnetic B-type star is surrounded by highly inhomogenous gaseous region. Rotating spots, arms, and
expanding regions have been observed, as well as the material infall and ejecta.
Farther from the star, one or two dust rings are present. The dust region is very clumpy, even large clouds have been detected.}\label{FSCMa_model}
\end{figure}

Figure~\ref{FSCMa_model} shows a~model of a~FS~CMa post-merger. The central star
is a~magnetic B-/early A-type star. The magnetic field is generated by mixing
during the merger \citep{Schneider20}. It may reach high values ($\sim$~6.2~kG
in IRAS 17449+2320, \citealt{IRAS_AA_mag}) that are comparable to the strongest
fields in Ap stars. Such a strong magnetic field slows down the stellar rotation
rapidly \citep{Schneider20} leading to the observed low rotation velocities
\citep[review in][]{hab}. The central star is probably a~pulsator, where 
the resonance of individual pulsating modes
may play a~role. The star undergoes episodic material ejecta approximately once a~month. The geometrically extended gaseous disk is very inhomogeneous.
Rotating arms and spots, as well as expanding layers, have been observed. The expanding layers slow down in some cases, even some material falls back onto the star \citep{Blanka}. Behind this gaseous disk, one or two dust rings are present. Dust regions are very clumpy, even large dust clouds were detected around some stars \citep{deWinter97}. The composition of these two rings may be different. Closer to the star, the graphite grains are present, whereas the outer region is composed of silicates \citep{Gauba03, Varga19}. The object has an extended low-density ``corona'', probably due to the
very strong magnetic field \citep{Alicia}. Seeing this object from the equator, the strong absorption of iron-group elements appears in the UV spectrum, so-called ``iron curtain''. The flux may be reduced by approximately an
order of magnitude compared to the classical B-type star \citep{Ivan-Pariz}. The absorption lines of the lower excitation stages also appear in the spectrum, because the disk behaves as a~``pseudo-atmosphere''. In this case, lines that should never be detected in B-type stars may be created. The most puzzling and important example is the Li I line resonance doublet \citep{lithium_conference}. The UV excess instead of the UV depression may be observed from the polar regions \citep{Bergner90,IRAS_AA_mag}.

The UV radiation is one of the key factors driving the variability in the
visible and IR regions. The absorbed UV energy is redistributed to the longer
wavelengths affecting not only the continuum, but also spectral lines. For
example, one of the related processes is connected with the overpopulation
of upper levels. The follow-up cascade of radiative deexcitation
leads to the appearance of the emission lines in the visible and IR. 
Without the knowledge of the behaviour of the UV radiation, it is hard to
properly decode the information hidden in the spectral lines. Therefore, UV
photometry is one of the essential techniques for the study such complex
objects.

A precise description of the short-term variability on the scale from
minutes to hours allows the search for pulsation modes of the central star. The
interference of the pulsation modes, leading to the so-called beating phenomenon, may be
responsible for ejecting matter in these stars. The determination of the
pulsation modes is especially important in FS~CMa post-mergers since these
objects are still out of the thermal equilibrium as shown by
simulations  of \citet{Schneider20} and position on the Hertzsprung–Russell
diagram \citep{Miroshnichenko_2017_HRD}. On-time scales around seven
hours, disk oscillations may be detected; similar to the disk flickering
observed in cataclysmic variables \citep{Bruch21} or symbiotic binaries
(Sect.~\ref{symbiotics}). To distinguish the disk oscillation from the stellar
pulsations, two bandpasses are necessary since the flickering amplitude is 
wavelength-dependent.

The photometry on longer time scales in two bands allows one to follow the
molecule and dust formation. Such an event has already been detected in MWC~349
\citep{White85}. Since the matter in the disk very efficiently absorbs the UV
radiation, a~raw guess of the angle of view is very easy to determine based on
the detected UV excess/depression. The UV properties and its variability are one
of the important ways to classify such a heterogenous group as FS CMa stars,
which is the necessary requirement for the search of new post-mergers. 

Even if only one intermediate-mass merger has been found yet in the phase when
the envelope started to be transparent, these post-mergers definitely deserve
more attention. As they represent the most frequent channel of mergers, their
contribution to the enrichment of interstellar matter may be significant.  They
may also contribute to the enrichment of intergalactic matter since the
simulations of \cite{Nela-Nbody} indicate that the merger may occur outside the
home galaxy. The position of FS~CMa stars in the Hertzsprung-Russell diagram
\citep{Miroshnichenko_2017_HRD}, around the terminal-age-main-sequence, with the
combination of the simulations of the merger event and evolution
\citep{Schneider20} shows that FS~CMa post-mergers are still out of thermal
equilibrium. These objects offer an attractive opportunity to study and
test stellar structure and evolution models. Last, but not least, less massive
FS~CMa post-mergers may be progenitors of magnetic Ap stars. Some of them may
offer an explanation for appearance of strong magnetic field in young
white dwarfs.  

The requirements for the mission parameters depend on the scientific goal.
While for the search for UV excess or deficiency one UV-band is sufficient,
study of variability typically requires a combination of far-UV and near-UV
bands. For most of the goals SNR of about 50--200 should be sufficient, but the
study of short-term variability requires an SNR of up to 1000 and a cadence of
10\,s.

\section{Star Clusters and ISM}%
\label{sec:clusters_ism}

\subsection{Photometric study of star clusters}

Star clusters, especially young ones, are relatively unexplored in the UV\@.
The understanding of the UV-bright stars and the populations they reside in --
clusters and stellar associations -- can be applied to study these objects in
other galaxies.  As we are only able to resolve the stellar populations in the
relatively nearby galaxies (up to a few Mpc away), the understanding of the
local clusters and associations dominated by UV-bright stars
\citep{2019AJ....158...35S} is crucial for studying these objects at 
extragalactic and cosmological distances, where they cannot be resolved.  These
distant populations are frequently metal-poor, so studying local (and those in
the Magellanic Clouds) low-metallicity clusters and associations will be
instrumental in testing of isochrones in the low-metallicity populations \citep{1995ApJ...454..767C}.

Many interesting types of stars (e.g.\ variables and binaries) can also be
investigated in star clusters.  In fact, it is preferable to investigate members
of star clusters because we have additional knowledge about their reddening,
distance, age, and metallicity. There are several types of unique objects that
could be reliably identified only in star clusters, mainly due to their
(unusual) position in the cluster colour-magnitude diagrams, where they pose as
outliers that deviate from isochrones.  Blue stragglers and extreme
horizontal branch stars are good examples of such objects.  Due to their
scarcity, they are not well understood, and the evolutionary pathways that lead
to their formation remain in doubt \citep{2011A&A...529A..60M}.  However, due to
their high temperatures when compared to the turnoff point, they emit a copious
amount of UV radiation, especially if the studied cluster is younger than about
1~Gyr.  A UV mission that gathers photometry of these objects would provide
helpful data that would supplement existing datasets and allow us to answer some
open questions regarding these objects. For example, these include the formation
scenarios for blue stragglers or extreme horizontal branch stars
\citep{2021MNRAS.507.1699J} and the role of binarity in the observed lack of
white dwarfs in open clusters \citep{2011Ap&SS.335..161B,2022arXiv220713992A}.

When we look at a colour-magnitude diagram of a star cluster, we can usually
identify the main-sequence turnoff (MSTO) point as the position of the bluest
(lowest colour index) main-sequence star in the diagram.  Since stars located
above this point in the colour-magnitude diagram evolve away from the main
sequence, astrophysicists use MSTOs to estimate the ages of star clusters.  
Because stellar rotation alters photometric indices and stars rotate with
different velocities, star clusters show so-called extended main-sequence
turnoff (eMSTO) \citep{2019NatAs...3...76L}. It has been suggested that some
clusters have experienced prolonged star formation and that age variation,
together with stellar rotation, is responsible for the eMSTOs
\citep{2017ApJ...846...22G}.  This effect has not been studied in UV so far.
Since the effects of interstellar and differential reddening can be readily
identified in UV
\citep{1979PASP...91..642T,2014A&A...569A...4D,2023MNRAS.522..367L},
investigating eMSTO in this part of the spectrum should help us to explain this
effect.  

The observations from \emph{Swift} UVOT have shown that the isochrones 
constructed using UV bands do not match the photometry of the red giants and
late-type main-sequence stars properly \citep{2020Ap&SS.365...95S}.  A possible
explanation is that the atmospheric models are incomplete in their treatment of
the UV absorption and emission lines originating due to circumstellar envelopes
and chromospheric activity (Sect.~\ref{kapstarchrom}). With the UV
mission,  the shortcomings of the current models could be tested.

\begin{figure}
    \centering
    \includegraphics[width=0.7\columnwidth]{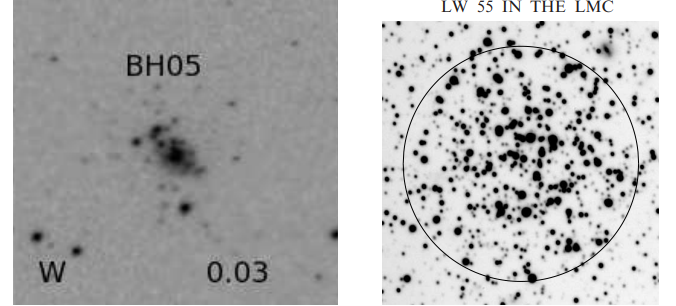}
    \caption{Extra-galactic clusters 
in Andromeda Galaxy (left) and in the Large Magellanic Cloud (right).  Images 
were taken from \citet{2009AJ....137...94C} and \citet{2003AJ....126..237K}.}
    \label{fig:gal}
\end{figure}

The use of colour-magnitude diagrams in the determination of the cluster
parameters is quite limited in the case of very distant Galactic and 
extragalactic clusters (for example, see Fig.~\ref{fig:gal}), where resolving
individual members is usually not possible \citep{2021arXiv210707230P}.
These problems can be overcome by using integrated photometry. Using nearby and
resolved star clusters, one can calibrate integrated photometry in terms of age,
reddening, and metallicity \citep{2002A&A...388..158L}.  With libraries based on
population synthesis codes, it is also possible to determine the total mass and
the fraction of red giants of star clusters \citep{2008BaltA..17..337B}.  This
method is well established, but it has not yet been applied to the UV region.

The photometric measurements of individual members of clusters in our Galaxy
requires a reasonably narrow point spread function to distinguish between
the individual stars. In general, we do not expect to reach distances
larger than about 3\,kpc, and anticipate more strict limitations within the
extinction-dominated Galactic disk. Integrated photometry of star clusters can
be conducted for local and extragalactic targets. For both objectives, the
main bandpass should mostly cover the 200--300~nm region. A secondary band
covering wavelengths unreachable from the ground ($ < 200$~nm) would be
beneficial. The FoV requirement is set by the size of the closest cluster
included in such a study.

\subsection{Deriving interstellar extinction in UV}

Interstellar dust is an important component of the interstellar medium (ISM).
Properties of the constituent particles are inferred from observations by
studying extinction curves together with polarisation curves, scattered light,
dust emission in the continuum, and absorption in bands
\citep{2003ARA&A..41..241D}.  Although an average extinction law is often
assumed when deriving extinction-corrected magnitudes of stars, 
looking in different lines of sight shows variations in the shape of the extinction curve that are connected to changes in the properties of dust \citep{1993A&A...274..439J}.

\begin{figure}
    \centering
    \includegraphics[width=0.7\columnwidth]{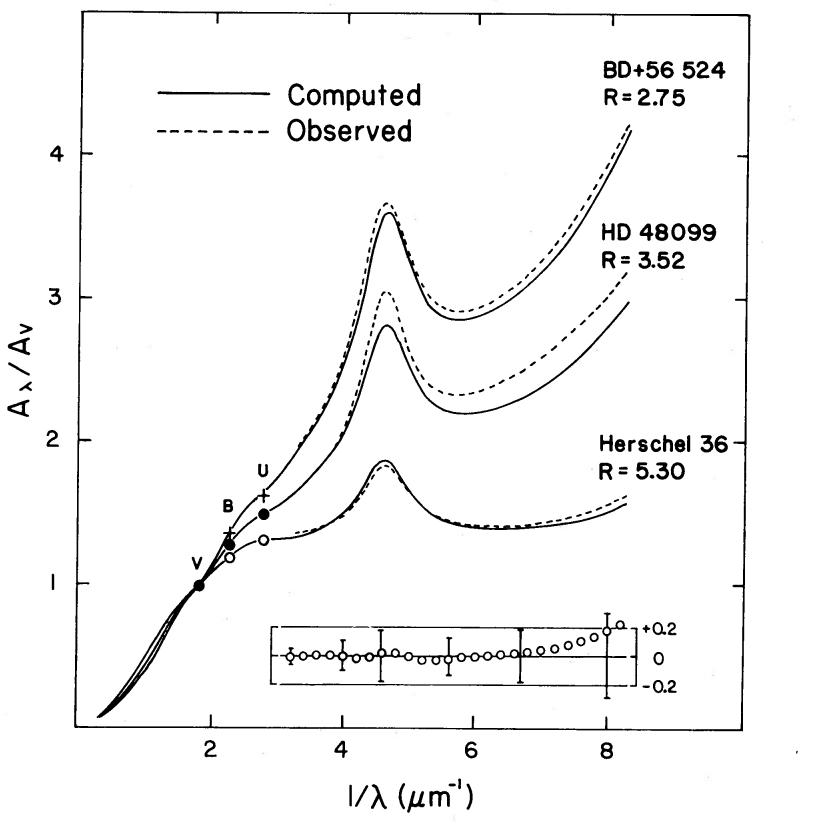}
    \caption{Extinction curves for
    different lines of sight \citep
    {1989ApJ...345..245C}.}
    \label{fig:ext}
\end{figure}

An extinction curve shows how the attenuation of light (measured, for example,
by extinction $A_{\lambda}$) depends on wavelength (Fig.~\ref{fig:ext}).  This
curve is affected not only by the chemical composition of the dust (mostly
silicates, graphite, ices, and polycyclic aromatic hydrocarbons) but also by the
sizes of the individual particles, ranging from a few microns down to a couple
of nanometres \citep[e.g.,][]{2001ApJ...548..296W}.  The strongest feature of
the extinction curve observed in near-UV is the extinction bump located at
around 217.5~nm \citep{1965ApJ...142.1683S}.

Based on IR photometric data, \citet{2005ApJ...623..897L} found that many
high-latitude clouds in our Galaxy show signs of an increase in the number
of smaller particles when compared with the diffuse ISM of the Galactic disk.
Variation in the size distribution of dust grains should affect not only the
strength of the UV extinction bump, but also the general steepness of the
extinction curve in UV\@.  This was explored by \citet{2021ApJS..256...46S}, who
made use of the UV data produced by the \emph{GALEX} mission and found that such
a situation mostly occurs in clouds with a relatively small amount of extinction
in the visible part of the spectrum.

A dedicated UV photometric survey would improve our understanding of this
effect.  Additionally, a good choice of photometric bands can help us to
distinguish between the UV continuum extinction and the strength of the bump at
217.5~nm, allowing us to study the variations in the dust grain populations in
better detail.  This is important for different studies of the ISM that focus on
variations in extinction curves, the evolution of dust grain populations, and
the formation of molecules.

To maximize the scientific output, it would be ideal to fit as many targets in
one field as possible -- for this reason, it would be ideal to have a FoV larger
than $1^\circ$.  Magnitude errors should be smaller than 0.1~mag, ideally 
as small as 0.01~mag. While the survey can work with almost any bandpass
in the 200--300~nm region, it would be preferred to make use of a band that is
focused on the wavelengths between 220~nm and 280~nm, which should best help to
capture the information about the profile of the UV extinction bump (when
combined with the available magnitudes at longer wavelengths).  A second band
centred at slightly longer ($\sim 340$~nm) or shorter ($\sim 160$~nm)
wavelengths would be beneficial to this study. The far-UV band is preferred
because it provides more information.

\subsection{Getting information about the ISM from UV bow shocks}

\begin{figure}
    \centering
    \includegraphics[width=0.95\columnwidth]{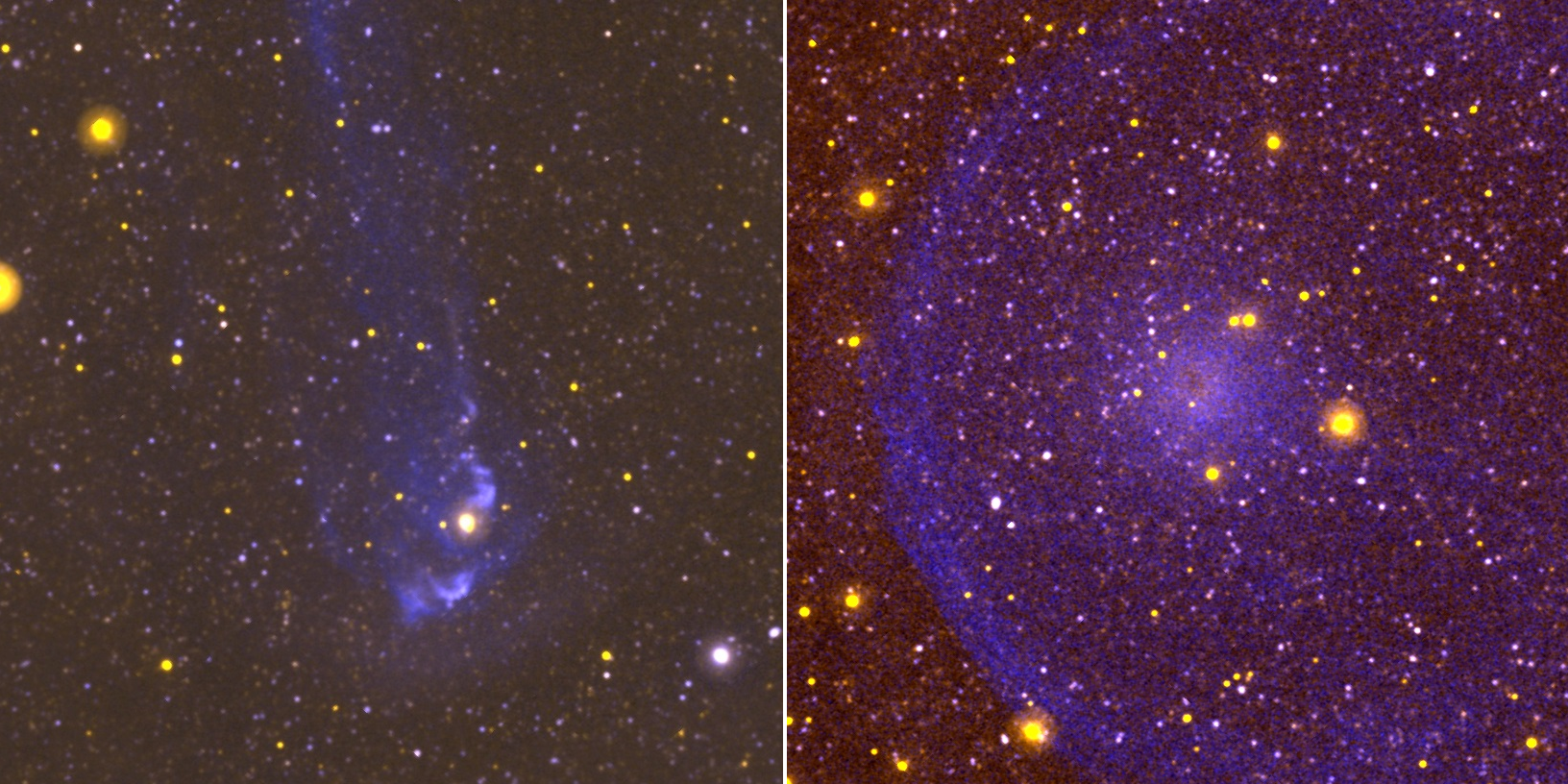}
    \caption{Composite UV images of bow shocks in the vicinity of
    Mira (left) and CW~Leo (right) taken by \emph{GALEX}.  Downloaded
    from \url{https://galex.stsci.edu/GR6/}.}
    \label{fig:bow}
\end{figure}

There are multiple events that can lead to the formation of shock waves in the
ISM.  One of them is the movement of a star with a strong stellar wind and a
high velocity with respect to the local ISM\@
\citep[e.g.][]{1988ApJ...329L..93V}.  The resulting bow-shaped shocks (or bow
shocks) can be observed at various wavelengths, most notably in the infrared
\citep {2001ApJ...554..778L,2007ApJ...657..810D}, optical \citep
{2005A&A...439..183B}, and at the higher energies as well. In a few cases
UV-emitting bow shocks have been found, for example, around Betelgeuse
\citep{2012A&A...548A.113D}, Mira \citep{2007Natur.448..780M}, and CW~Leo
\citep{2010A&A...518L.141L} (see Fig.~\ref{fig:bow}).

However, the UV emission of bow shocks is not well-constrained since their
observation in this spectral region is quite challenging.  Firstly, their
driving stars are often strong UV emitters, making it difficult to resolve the
rather faint, diffuse, and wispy emission of a bow shock in their proximity.
Moreover, some missions such as \emph{GALEX} intentionally avoided many bright
UV stars, in order to protect its detectors \citep{2015ApJ...800..132C}.  Some
bow shocks from
\citet{2010A&A...519A..33G,2011A&A...525A..17G,2016ApJS..227...18K} were
serendipitously observed by \emph{Swift} UVOT, but so far no clear bow shock
emission associated with O and B stars has been detected in UV.

Detecting these structures in UV would provide valuable data to complement
the observations more routinely done in other bandpasses. One of the main
focuses would be to put constraints on the physical processes that dominate in
the bow shock regions. Several processes may influence the bow shock emission in
UV, including absorption \citep{2012A&A...548A.113D} and scattering by dust
particles, emission lines resulting from collisional excitation
\citep{2007Natur.448..780M}, and to a smaller degree also, inverse Compton
scattering \citep{2018ApJ...864...19D}.

The bow shocks can be detected from frame images focused on regions surrounding
asymptotic giant branch (AGB) stars, and O and B stars. Since the
angular radius of a bow shock depends on the distance (and other parameters), a
reasonable FoV ($\sim 1^\circ$) is required.  The angular resolution
should be sufficiently high to be able to map the structure of the bow
shocks. The magnitude limit is mostly influenced by the observability of O
and B stars.  While most of the AGB stars are expected to have $m_{\mathrm{UV}}
> 9$~mag, O-type stars of interest have $m_{\mathrm{UV}} > 4$~mag (if we focus
on objects beyond 1~kpc, $m_{\mathrm{UV}} > 6$~mag).  For this reason, it could
be possible to allow having the observed objects saturated in some cases. To be
able to identify bow shocks, the signal-to-noise ratio has to be higher than 3.
Although an AGB star at 3~kpc could be as bright as $m_{\mathrm{UV}} = 16$~mag,
the situation changes when the presence of the ISM is assumed, which would shift
the lower limit to $m_{\mathrm{UV}} \geq 20$~mag.

\emph{GALEX} images suggest that UV bow shocks appear brighter at shorter
wavelengths.  While some results are already expected for a near-UV
region, a far-UV band should enable the identification of more bow shocks.

\section{Exoplanets}%
\label{sec:exoplanets}

In the past three decades, the characterization of thousands of extra-solar
planets allowed us to make further steps towards the understanding of the origin
of life on Earth and the possible discovery of signs of life outside our Solar
System.  In terms of methodology, numerous techniques have proven to be useful
for confirming the presence of a planet orbiting a distant star, such as pulsar
timing variations \citep{wolszczan1992}, radial velocity measurements
\citep{mayor1995} or direct imaging \citep[see, e.g.][]{lagrange2020}. However,
the vast majority of known systems were discovered and investigated via
photometric measurements of planetary transits, i.e.\ when the planet blocks a
small but well-detectable amount of stellar light in coincidence with its
orbital period \citep{henry2000}.  The dominance of transiting planets within
the family of known planets is mostly due to the significant advantage of
space-borne photometry: it yields not just precise and accurate time series, but
the lack of diurnal variations allows for the extension of the time domain
compared to ground-based telescopes.  Therefore, space telescopes play an
important role in the analysis of these systems.

\subsection{Transiting exoplanetary systems}

While radial velocity measurements are essential to derive the mass ratios
and orbital characteristics of exoplanetary systems, the information provided by
photometric transits (and occultations) significantly increases our knowledge by
revealing the size and, therefore, the density of the planets.  The first such
confirmation of the presence of transits happened in the case of the planetary
system HD~209458(b) \citep{charbonneau2000} while, a few years later, the
transit method was applied for the discovery of OGLE-TR-56 \citep{konacki2003}.
In addition, precise time series associated with individual transits not only
reveal a more precise orbital solution in terms of parameters such as orbital
period or inclination, but also play a key role in understanding the host star
\citep{seager2003}.  Although these key discoveries and confirmation of transits
were also made by space-borne photometry \citep[see, e.g.][]{brown2001},
dedicated space missions such as \textit{CoRoT} \citep{barge2008},
\textit{Kepler} \citep{borucki2010} or \textit{TESS} \citep{ricker2015},
literally increased the number of known such systems by thousands.

Similarly to our Solar System, perturbations occur between different
planets orbiting the same star. By their nature, such perturbations 
are highly dependent on the mass of the planets. One of the most easily and
precisely observable effect of such perturbations is the variation
in the orbital phase of the planetary companions. Periodic or systematic
variations in the orbital phases cause measurable changes in the transit timings.
Observations of these transit timing variations allow us to
determine the masses of the planets \citep{holman2010}. In addition,
perturbations having an intrinsically secular nature also yield observable
changes in the duration of the transit. For instance, the oblateness of the host
star causes the planetary orbit to precess, implying a slight but detectable
systematic shift in the duration of the transits \citep{szabo2012}. Such
measurements therefore reveal intrinsic parameters of the host star, like the
non-spherical components of its gravitational field. Taking into account
the variations in the parameters of exoplanetary systems, such as transit timing
variations or transit duration variations, it is also possible to reveal the
presence of another companions that would otherwise remain undetected. This is
particularly relevant for companions on orbits with lower inclinations, on which
the exoplanets do not transit the star, or have so small masses that they do not
provide sufficiently high amplitude signals in radial velocities.

The shape of transit light curves are also determined by the stellar limb
darkening \citep[see, e.g.][]{brown2001}. Quantitatively, the limb darkening
depends on both the central wavelength of the observations and the stellar
atmospheric parameters \citep{claret2004}, allowing further
refine the host star parameters (such as the independent measurements of the
surface temperature or spectral type). Limb darkening is more prominent on
shorter wavelengths. Precise space-borne observations are important to manage
the correlations between the planetary radius and the derived limb darkening
parameters \citep{pal2008lcdiff}. In addition, the presence of an atmosphere can
yield a wavelength-dependent planetary radius \citep[see, e.g.][for an
introduction]{deming2019}. Stellar spots also affect transiting light curves
\citep{nutzman2011,oshagh2013}, and the shorter the wavelength used for
observation, the greater the contrast the stellar spots have, allowing a clear
distinction of stellar spots from instrument noise. 

The study of exoplanetary transits is relatively demanding for the mission
parameters. The requested SNR of about 500 within one-hour exposition
could be accordingly modified if faster cadence of about 1\,min is needed.

\subsection{Activity, flares, and star-planet interaction}\label{secexoact}

By nature, flares, and the corresponding physical processes (such as coronal
mass ejections) associated with a star harbouring a planetary system affect
habitability: the higher the flaring rate, the more energetic the ionizing
radiation threatening the potential life on a planet.  On the other hand, having
no flaring activity on a host star might not provide sufficient input for
triggering the formation of fundamental molecules essential to life.
Main-sequence stars with a small mass (i.e.\ mostly M dwarfs or K dwarfs) are
the focus of research in the field of exoplanets due to the easier
detection of planetary companions \citep[see, e.g.][and the references
therein]{irwin2011}.  These objects tend to have an enormous amount of flaring
activity, as shown by recent individual \citep{vida2019} and statistical
\citep{seli2021} analyses.  While observatories like \textit{TESS}, operating at
long wavelengths, provide reliable constraints in a statistical manner for
a large number of objects \citep[see also][]{seli2021}, flaring activity is the
most prominent at short wavelengths, such as the UV band
(Sect.~\ref{kapstarchrom}).

It is well known that stars strongly affect their close cool companions via
gravity and irradiation (see, e.g.\ Sect.~\ref{sec:binaries}). It is less
obvious that planets also affect their host stars. Late-type stars store a
significant amount of energy in their surface magnetic fields.  Giant planets on
close orbits may have their own magnetic fields and magnetospheres that may
interact with the stellar magnetic fields \citep[see][for a review]{vidotto20}.
As the planet moves over the surface of the star, it may trigger reconnection
events that will follow the planet.  This may heat the upper atmosphere of the
star and cause variability with a period equal to the orbital period of the
planet. Apart from that, the stellar wind may interact with the planetary
magnetic fields, creating a bow shock and an asymmetry in the circumplanetary
material. This was detected as an early ingress during the transit of
WASP-12b in UV with the Hubble Space Telescope ({\it HST})
\citep{fossati10}.  Detection of this kind of interaction is important because
it would provide information on the magnetic field of the planet, which is
almost impossible to obtain by other methods.  The magnetic field of the planet
is crucial for the assessment of the habitability of the planet. 

UV photometry is particularly well suited to study such planet-star 
interactions.  This is because UV flux originates from the upper atmosphere. UV
flux of late-type stars constitutes only a small fraction of their total energy
output. Hence, the UV band is very sensitive to such high-energy phenomena
even if the flux in the visual band remains almost constant.  Study
of planet-star interactions will require monitoring of about a dozen stars with
massive close-in planets for weeks.  This task is too expensive for large
telescopes such as \textit{HST} and small-scale missions such as \emph{QUVIK}
can make a significant contribution in this field.

\subsection{Exo-atmospheres at short wavelengths -- the border between life and death}

UV photometric mission will offer a unique window into the characterization of
exoplanetary atmospheres. The energetic radiation heats the upper layers of
planetary atmospheres, making the atmospheric material easier to escape.
In the UV part of the electromagnetic spectrum, the planetary diameter appears
larger, and thus transits are deeper than at longer wavelengths.  This allows us
to study the evaporation of atmospheres and atmospheric escape, which is
particularly strong in close-in planets around M-dwarf stars \citep[for example,
GJ436~b,][]{2015Natur.522..459E, 2017A&A...605L...7L}.

There is a well-known gap in the radius distribution of exoplanets ($1.6-2\,
R_{\rm Earth}$) and a complete lack of Neptun-sized exoplanets with orbital
periods shorter than 3 days \citep[e.g.][]{2012ApJS..201...15H,
2014ApJ...787...47S, 2017AJ....154..109F, 2018MNRAS.479.4786V}.  This is
explained as a result of exoatmospheric erosion and has been predicted by many
different escape models \citep[e.g.][]{2012ApJ...761...59L,
2018MNRAS.476..759G}.  Observations by small-sized UV telescopes will help us to
understand the atmospheric escape better and give us a better idea of the
capability of the planets to keep their atmospheres.  This is especially
important in studying exoplanets in habitable zones that could potentially host
extraterrestrial life.

\section{Transiting dusty objects: from exoasteroids to dusty discs}%
\label{sec:transiting_dusty_objects}

Dust is known to scatter and absorb UV radiation very efficiently.  At the same time, hot stars are strong emitters of UV radiation.  Consequently, UV telescopes are ideal for detecting and studying transits of dust clouds in front of sufficiently bright and hot stars.  
While the nature and origin of stars and dust are usually specific to a given type of star, the dust obscuration process is universal.

\subsection{Disintegrating exoplanets}

Most of the known exoplanets transit their host stars and feature stable and symmetric transits.  A decade ago, a new class of objects was discovered \citep{rappaport12} that exhibit asymmetric, strongly variable but periodic transit-like signals.  
Sometimes they feature a strange pre-transit brightening.  Kepler-1520b became a prototype of such objects.  On the other hand, some transits, such as those of K2-22b \citep{sanchis15}, look almost the opposite, with a post-transit brightening.  Periods are shorter than one day, and transit durations are approximately 1 hour.

It turned out that such transits are caused by disintegrating exoplanets that
orbit very close to their host stars.  The rocky surface of the planet
evaporates and forms an opaque dusty tail.  It is this tail that is causing the
transit, while the planet itself is usually too small to be detected.  The tail
(dust cloud) is very sensitive to various processes, such as
radiative/gravitational accelerations, dust evaporation, or condensation. It
efficiently scatters the radiation in the forward direction, causing the
above-mentioned pre- or post-transit brightenings, depending on whether the tail is trailing or leading the planet.  Such planets might have lost a significant fraction of their mass, exposing their naked cores \citep{perez13}.  Consequently, the properties of the dust tail reflect the bulk (not only the surface) chemical composition of the planet and provide a unique tool for probing the interiors of the planet.  

There are half a dozen of such objects. Generally, their transits would be difficult to detect in UV since stars are typically of later spectral types (F-M), faint ($V$\,=\,16 mag), and the transit depth is often below 1\%. 
However, there are related objects that are either brighter and/or the transits
are deeper, and these are mentioned in Sect.~\ref{related}.  Ongoing and planned space missions such as \textit{TESS} \citep{ricker2015} and \textit{PLATO} \citep{rauer14} may discover brighter objects of this kind.

\subsection{Exoasteroids}

\begin{figure}
  \centering
  \includegraphics[height=4.8cm]{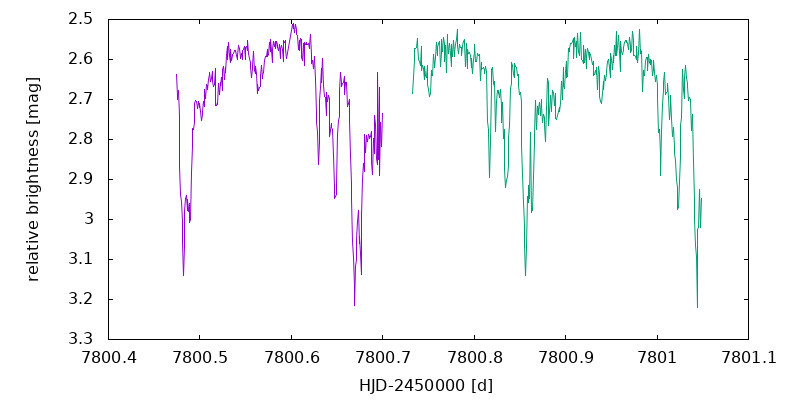}
  \caption[]{Light curve of WD1145 in the optical region observed in 2017 (Maliuk et al., in preparation). Everything deeper than an uncertainty (0.02 mag) are transits.}
  \label{fig:1}
\end{figure}

Since the discovery of disintegrating exoplanets, even smaller (minor) bodies have been detected to orbit and transit other stars -- ``exoasteroids''.  White dwarfs offer an advantage in detecting such objects.  Their radii are 100 times smaller than the Sun, which enables the detection of relatively small objects.  The first known exoasteroids were discovered orbiting a white dwarf named WD1145+017 \citep[WD1145]{vanderburg15}.  They have orbital periods of about 4--5 hours, and transits in the optical region may be 40--60\% deep.  An example of the light curve from the 2017 season is shown in Fig. \ref{fig:1}.  One can see numerous dust clouds associated with asteroids passing in front of the star, causing eclipses.  This behaviour is in many respects similar to the disintegrating exoplanets mentioned above.  The nature and variability of such transits call for simultaneous multi-wavelength observations to put constraints on the chemical composition and properties of dust.
Surprisingly, UV observations unravelled that transits at these wavelengths are significantly shallower than those in the optical region \citep{hallakoun17,xu19}.  The reason for that is still an open question, but it may be a combination of two factors: an underlying absorption by a co-planar gas disk and a forward scattering on large dust particles \citep{xu19,budaj22}.

The discovery of such exoasteroids solves a missing chain in our understanding
of the chemical composition of white dwarfs.  Many white dwarfs show heavy
elements in their spectra.  However, these elements should have quickly sunk due
to the strong gravity of the star \citep{paquette86}. WD1145 unravelled the
expected chain of events before our eyes for the first time
\citep{debes02}: stars have planetary systems; orbits of bodies are
disturbed by mass loss during the AGB phase; they get close to the WD (a remnant
of the AGB star) and are disrupted into pieces, forming dust clouds; dust turns into a gaseous disk as it approaches the hot star; and, finally, gas accretes onto the surface of the star, contaminating its chemical composition.  Consequently, by measuring the chemical composition of the star or its gaseous disk, we also infer the chemical composition of the disrupted body.  For example, it appears that WD1145 recently encountered and accreted a body of Earth-like composition \citep{xu16,fortin20,budaj22}.  

The characteristic time scale and the amplitude of variability due to
transits of exoasteroids are 10~minutes and 0.1~mag, respectively. Required
SNR and cadence are 50 and 2 min, respectively. An advantage of such white
dwarfs is that they are hot (characteristic spectral type is DB) but
relatively faint ($U = 17$--$18$~mag).  

Since the discovery of WD1145, there has been increasing evidence that
asteroids, planets or their remnants are also orbiting other white dwarfs:
ZTF J013906.17+524536.89 \citep{vanderbosch20}, SDSS J122859.93+104032.9
\citep{manser19}, and WD~J091405.30+191412.25 \citep{gansicke19}.  Most
recently, a Jupiter size planet was directly detected to transit WD~1856+534
\citep{vanderburg20}.  However, there are still only a handful of such objects,
and little is known about their incidence, orbits, masses, chemical composition, evolution, origin, and other properties.  Increasing these numbers is a crucial next step in answering all these questions.  A small UV telescope may be used to search for such variability by targeted observations of selected white dwarfs, by post-processing of full-frame images, or for follow-up observations of new objects detected by other missions (\textit{TESS}, \textit{PLATO}).

UV photometry will verify the nature of the event.   
Dust clouds will have variable asymmetric transits of different depths at different wavelengths.  Based on the shape of the transit, it will be possible to put constraints on the shape of the dust cloud (e.g.\ trailing/leading tail, disk).

Dust properties, such as particle sizes and chemical composition, are another
important question that requires multi-wavelength observations.  UV photometry
in two bandpasses will constrain these properties.  However, since the
depths of transits are variable, they will have to be coordinated with
additional simultaneous observations in the optical and IR regions.  Apart from
that, it will be possible to put constraints on the total amount of dust and its
production rate, which will constraint the masses and lifetimes of the dust-producing bodies.

An advantage of the UV region is that it is most sensitive to small dust
particles ($<0.1\,\mu\text{m}$) that are difficult to detect at longer
wavelengths.  Consequently, transits may be deeper in the UV region and may not
even be seen at other wavelengths.

\subsubsection{Related objects}%
\label{related}

There are stars that are brighter than the above-mentioned white dwarfs and are
related to the topic.  Boyajian's star \citep{boyajian16} is a famous example.
It is an F-type and \textit{V}\,=\,12 mag star.  Its Kepler light curve shows a
few strong dimming events that are 10--20\% deep, asymmetric, and aperiodic.
Ground-based follow-up observations revealed multiple dips of about 1\%
\citep{boyajian18}.  It is a `normal' star on the main sequence, not a young
star. Multi-wavelength observations indicate that this variability is compatible
with a dust extinction caused by a swarm of comets \citep{bodman16} or dust
clouds associated with more massive bodies such as asteroids \citep{neslusan17}.
\textit{TESS} and \textit{PLATO} missions should discover more objects of
this kind.  Characteristic timescale and amplitude of variability are 1 day and
0.01--0.1~mag, respectively. This translates into SNR of about 200 and the
cadence $<1$ day that is required.

\subsection{Eclipses by dusty disks and rings}

There are also other dusty objects much larger than the aforementioned
minor bodies that transit stars. Some long-period eclipsing binary stars have
very peculiar, variable, and asymmetric eclipses such as $\varepsilon$~Aur 
(spectral type F0, $V = 2.9$~mag).  It is a bright binary star with an orbital
period of 27~yr and deep eclipses lasting for almost 2~yr \citep{ludendorff03}.
The fascinating story of this star dates back 200 years and is full of
challenges, twists, and turns.  The binary consists of a giant F star and a
mysterious huge invisible component. Only multi-wavelength and interferometric
observations during the latest eclipse proved the nature of the companion, which
is a cool dusty disk \citep{kloppenborg10}. Nevertheless, it was widely
believed that the disk was inclined to the orbital plane and that the light from
the F star penetrated through the central hole in the disk, causing mid-eclipse
brightening \citep{wilson71,carroll91}.  Surprisingly, it was shown that a
co-planar but flared disk geometry (with a vertical thickness increasing
with distance) and forward scattering could explain the strange eclipse more
naturally \citep{chadima11epsaur,budaj11epsaur}.  Such objects are rare, and we
know of only a few long-period binaries with such dusty disks. However, their
numbers are growing, and most of them were discovered only during the last
decade, thanks to modern surveys and their combination with digitised archival
data dating back more than a century.  The nature/origin of their variability,
central stars, disks, and properties (shape, dynamics, dust particle
sizes, and chemical composition) are still a matter of debate.  The
characteristic duration and amplitude of variability are 1~yr and 0.1--1~mag.,
respectively. The SNR and the cadence required are about 100 and 1--10
days, respectively.

From the shape, duration, and variability of the eclipse, it will be possible to
identify disks, rings, holes, gaps, clouds, or warps if present.  Some of these
structures may be associated with otherwise invisible small bodies orbiting
inside the disk, which is interesting from the point of view of the planet
formation \citep{kenworthy15}.  An advantage of the UV region is, again, that it
is more sensitive to dust, and it may detect smaller dust particles that are
invisible at longer wavelengths.  If the UV observations are combined with
observations at other wavelengths\footnote{For example ZTF data covering optical
region \citep{bellm2019} may be very useful.}, this will put constraints on the
dust particle size, its chemical composition, and the total mass of the dust.

Apart from that, observations in the UV region may help solve the main problem: what is inside the dusty disk? The light from this embedded object is heavily attenuated by the disk, which makes it difficult to detect the object.  However, if this is a hot object, the disk atmosphere will efficiently scatter its UV light towards the observer, and the object may be detected in UV\@.  During the eclipse, the main source of the light is suppressed, which further facilitates the detection of the hidden, hot object.  This can be used not only for dusty but also for gaseous disks.  UV photometry can be combined with spectroscopy, interferometry, and photometry at other wavelengths, to get a deeper insight into the structure of the whole system \citep{broz21}.  

Another similar class of objects on a completely different scale are possible transiting circum-planetary (exoplanetary) rings or exorings.  The presence of such rings has been recently hypothesised for some transiting exoplanets.  It is related to the problem that the radii of some exoplanets are so large that their average densities are of the order of 0.1~g/cm$^3$, which is an unexpectedly low value. The presence of exorings around these planets is an alternative solution of this problem \citep{akinsanmi20,piro20}.
UV and IR observations are of crucial importance for detecting such rings.  From optical observations alone, it is almost impossible to distinguish whether the transit is due to a seemingly large planet or a small planet with a dusty ring.  An example is the exoplanet HIP 41378 orbiting a bright (V=8.9 mag) F8-type star \citep{akinsanmi20}.  

There are also other fascinating events/stars in this category.  One of the
biggest and brightest stars in the sky, a red supergiant star Betelgeuse 
(spectral type M1\,Ia, $V = 0.0$~mag), recently dimmed unexpectedly by a factor
of 3, reaching a historical minimum in 2020 \citep{guinan20}.  An
increased UV continuum was observed prior to the dimming. The nature/origin of
the variability is an open question.  There were speculations about an imminent
supernova eruption. However, UV observations with HST indicated that it could be
due to a dust cloud passing in front of the star \citep{dupree20}.  Another
question arises about the origin and properties of the dust cloud.  It might
have originated from the star itself (see also Sect. \ref{betelgeuse}).  UV
observations would be valuable, since the IR observations did not detect any
significant change in brightness \citep{betnedim,jadbet}.  Characteristic
duration and amplitude of variability are 1 month and 1~mag. This translates
into SNR and cadence that are required of about 100 and 1 week,
respectively.

R Coronae Borealis (RCB) stars are another type of supergiants featuring
irregular, several magnitudes deep fading events that may last for months or
years. They are of spectral type F-G, hydrogen-poor, carbon-rich, and eclipses are most likely caused
by the dust condensation in the vicinity of the star, but the nature of the
stars and the events are not clear.

\subsection{Dippers}

Young dipper stars experience eclipses by dusty disks too. 
About 20--30\% of classical T~Tau stars are dippers 
\citep{2010A&A...519A..88A,2011ApJ...733...50M}.
Most of these young stars are of late spectral types (K--M) and feature dusty protoplanetary disks.  Such a disk may have a complicated structure with an inner disk that is misaligned with the outer disk.  Day-long eclipses are often observed in these stars.  It is believed that they are caused by dust from the inner disk that is nearly edge-on, causing an occultation of the star.  Consequently, the light curve is very sensitive to any dust clumps at the edges of the inner disk.  However, the nature of these dust clumps is unknown.  They may be due to warps in the inner disk, vortices, or clumps associated with planet formation \citep{ansdel16} or may be associated with magnetospheric accretion \citep{bodman17}.  Dips may be quasi-periodic or aperiodic.  The characteristic duration and the amplitude of the variability are 1 day and 0.1~mag, respectively. The required SNR is about 100 and a cadence of about a few hours.

There is another group of young stars called UXors after the prototype UX~Ori.
They are very similar to the above-mentioned dipper stars but are slightly more
massive, hotter, and are often associated with Herbig Ae/Be stars 
\citep{1991Ap&SS.186..283G}.  They feature an infrared excess indicating a dusty disk,
semi-regular dimmings on the time scales of months to years, and possible brightenings due to accretion or rotational modulation of variability due to spots.

Long-term multi-wavelength photometry will determine a possible periodicity of the events and put constraints on the geometry of the inner disk and dust clouds.  In the case that the dips are periodic, it will be possible to locate their source based on the period.  This may lead to the discovery of dust clouds or holes associated with planet formation.  UV observations will be quite sensitive to such structures.
They are also ideal for detecting hot spots on the surface of the star and 
determination of the rotational period of the star.  Such spots are associated with magnetospheric accretion, which, in turn, indicates the presence of magnetic fields.  This will put important constraints on the corotation and Alfv\'en radii that control the magnetospheric accretion and eventually shed more light on the origin of dips.

Typical dipper stars \citep{ansdel16} are K--M stars with \textit{J} magnitudes in the range 8--13~mag, which corresponds to \textit{U} magnitudes of about 14--19~mag.  Consequently, those dipper stars and UXors that are brighter and hotter should be observable with a small UV telescope.

Multi-wavelength observations will also put constraints on the dust properties and chemical composition.
Using two bandpasses in UV would help to constrain the temperature of the star from the slope of its continuum.  However, since most of these dipper stars are of late spectral types and faint, we do not expect to detect many of them in the far-UV band.

\section{Conclusions}

We have demonstrated the impact of a small-size two-band UV photometric satellite on the astrophysics of stars and stellar systems. We have shown that UV mission can improve our understanding of physics and evolution of virtually all types of stars, including hot and cool main-sequence stars, supergiants, and compact remnants such as neutron stars. UV photometry provides constraints for precise determination of binary parameters, probes the circumbinary environment, and leads to significantly better understanding of different evolutionary stages of binaries including stellar mergers. Most conveniently, stellar physics can be tested for members of star clusters taking advantage of the common distance of all its members. UV mission can provide crucial information on exoplanets, including their basic parameters, atmospheres, and especially their interaction with host stars. UV observations can probe also other parts of exoplanetary systems from exoasteroids to large dusty disks.

Ground-braking observations can be obtained already using a single-band UV mission. However, we have demonstrated that including a second band can tremendously increase the scientific output of the mission. The optimal combination includes far-UV and near-UV bands.

We have listed the requirements that the satellite should met to fulfill individual scientific goals. The required parameters include the limiting magnitudes, final signal to noise ratio, and cadence of observations.

\paragraph{Acknowledgement}
JBu was supported by VEGA 2/0031/22 and APVV-20-148 grants.
JBe acknowledges support from the German Science
Foundation (DFG) projects BU 777-17-1 and BE 7886/2-1.
\section*{Declarations}

\paragraph{Competing Interests} The authors declare no competing interests.

\bibliography{sn-bibliography}
\bibliographystyle{aa} 


\end{document}